\documentclass{article}

\usepackage{arxiv}

\usepackage[super]{nth}
\usepackage{caption}
\usepackage{subcaption}
 \usepackage{array,multirow,graphicx}
 \usepackage{float}
\usepackage{nameref,hyperref}

\usepackage{tabularx}
\usepackage[super]{nth}
\usepackage{adjustbox}
\usepackage{xcolor}

\title{FakeNewsLab: Experimental study on biases and pitfalls preventing us from distinguishing true from false news}

\author{
Giancarlo Ruffo \\
  Dipartimento di Scienze e Innovazione Tecnologica (DISIT) \\
  Università del Piemonte Orientale \\
  Alessandria, Italy\\
  \texttt{giancarlo.ruffo@uniupo.it} 
   \And
Alfonso Semeraro \\
  Dipartimento di Informatica\\
  Universit\`a degli Studi di Torino\\
  Torino, Italy \\
  \texttt{alfonso.semeraro@unito.it} 
}

\begin{document}

\maketitle

\begin{abstract}
Misinformation posting and spreading in Social Media is ignited by personal decisions on the truthfulness of news that may cause wide and deep cascades at a large scale in a fraction of minutes. When individuals are exposed to information, they usually take a few seconds to decide if the content (or the source) is reliable, and eventually to share it. Although the opportunity to verify the rumour is often just one click away, many users fail to make a correct evaluation. We studied this phenomenon with a web-based questionnaire that was compiled by 7,298 different volunteers, where the participants were asked to mark 20 news as true or false. Interestingly, false news is correctly identified more frequently than true news, but showing the full article instead of just the title, surprisingly, does not increase general accuracy. Also, displaying the original source of the news may contribute to mislead the user in some cases, while a genuine wisdom of the crowd can positively assist individuals’ ability to classify correctly. Finally, participants whose browsing activity suggests a parallel fact-checking activity, show better performance and declare themselves as young adults. This work highlights a series of pitfalls that can influence human annotators when building false news datasets, which in turn fuel the research on the automated fake news detection; furthermore, these findings challenge the common rationale of AI that suggest users to read the full article before re-sharing.
\end{abstract}



\section{Introduction}
\label{sec:introduction}

Information disorder~\cite{CouncilofEurope,UnescoHandbook}, a general term that includes different ways of describing how the information environment is polluted, is a reason of concern for many national and international institutions; in fact, incorrect beliefs propagated by malicious actors are supposed to have important consequences on politics, finance and public health. The European Commission, for instance, called out the mix of pervasive and polluted information that has been spread online during the COVID-19 pandemic as an ``infodemic''~\cite{EuropeanCommission}, which risks jeopardising the efficacy of interventions against the outbreak. In particular, social media users are continuously flooded by information shared or liked by accounts they follow, and news diffused are not necessarily controlled by editorial boards that in professional journalism are in charge of ruling out poor, inaccurate or unwanted reporting. However, the emergence of novel news consumption behaviours is not negligible: according to a research of the Pew Research Center~\cite{PewJournalism}, 23\% of United States citizens stated to get their information from social media often, while 53\% at least sometimes. Unfortunately, even if social media may actually increase media pluralism in principle, the novelty comes with some drawbacks, such as misinformation and disinformation that are spreading at an amazingly high rate and scale~\cite{Lazer1094,Allcott2017,Vosoughi1146,Grinberg374}. 

If the reliability of the source is an important issue to be addressed, the other side of the coin is individual's intentionality of spreading low quality information; in fact, although the distinction between misinformation and disinformation might seem blurry, the main difference between the two concepts lies in the individual's intention, since misinformation is false or inaccurate information deceiving its recipients that could spread it even unintentionally, while disinformation refers to a piece of manipulative information that has been diffused purposely to mislead and deceive. Focusing on misinformation, we may wonder how and why users can be deceived and pushed to share inaccurate information; the reasons for such behaviour is usually explained in terms of the inherent attractiveness of fake news, that are more novel and emotionally captivating than regular ones~\cite{Vosoughi1146}, the repeated exposure to the same information, amplified by filter bubbles and echo chambers~\cite{Flaxman2016FilterBubbles}, or the ways all of us are affected by many psycho-social and cognitive biases, e.g., the bandwagon effect~\cite{Nadeau1993}, confirmation bias~\cite{ConfirmationBias}, the vanishing hyper-correction effect~\cite{butler2011}, that apparently makes debunking a very difficult task to be achieved. However, much remains to be understood on how the single user decides that some news is true or false, how many individuals fact-check information before making such decision, how the interaction design of the Web site publishing the news can contribute in assisting or deceiving the user, and to which extent social influence and peer-pressure mechanisms can activate a person into believing something against all the evidences supporting the contrary.

Our main objective is to investigate the latent cognitive and device-induced processes that may prevent us from distinguishing true from false news. To this purpose, we created an artificial online news outlet where 7,298 volunteers were asked to review 20 different news, and to decide if they are true or false. The target of the experiment is manifold: we aim to assess if individuals make these kinds of decisions independently, or if the ``wisdom of the crowd'' may play a role; if decisions made in function of limited information (the headline and a short excerpt of the article) lead to better or worse results than choices made upon richer data (i.e., the full text, the source of the article); if users' better familiarity with the Web or other self provided information can be linked to better annotations. The results of the experiment and their relationships with related work are described in the rest of the paper.

We think that understanding the basic mechanisms that lead us to make such choices may help us to identify, and possibly fix, deceiving factors, to improve the design of forthcoming social media web-based platforms, and to increase the accuracy of fake news detection systems, that are traditionally trained on datasets where news has been previously annotated by humans as true or false, since human annotators' biases may leak into the corpora that fuel the research on fake news detection. Furthermore, such systems often rely on features such as the credibility attributed to the source of an article~\cite{yuan-etal-2020-early, Sitaula2020}, or how the news article is perceived by human readers~\cite{credbank, weak_social_sup, Tschiatschek2018}; we show in this work how these features can be misleading. Last, AI-generated messages suggest users online to read entirely an article before sharing it, a counter-measure to impulsive re-sharing that we show to be ineffective.

\section{Related Works}
\label{sec:relatedworks}

Telling which news is true or false is somehow a forced simplification of a multi-faceted problem, since we must encompass a substantial number of different bad information practices, ranging from fake news to propaganda, lies, conspiracies, rumours, hoaxes, hyper-partisan content, falsehoods or manipulated media. The information pollution is therefore not entirely made of totally fabricated contents, but rather of many shades of truth stretches. This complexity adds to the variety of styles and reputation of the writing sources, making it hard to tell misleading contents from accurate news. Given that users' time and attention are limited~\cite{qiu2017limited}, and an extensive fact-checking and source comparison is impossible on large scale, we must rely on heuristics to quickly evaluate a news piece even before reading it. Karnowski et al.~\cite{KARNOWSKI201742} showed that users online decide on the credibility of a news article by evaluating contextual clues. An important factor is the source of the article, as pointed out by Sterrett et al.~\cite{Sterrett2019}, who found that both the source (i.e., the website an article is taken from) and the author of the social media post that shares the article have a heavy impact on readers' judgement of the news. Also Pennycook et al.~\cite{Pennycook2521} investigated the role of the website that news is taken from, finding that most people is capable of telling a mainstream news website from a low-credibility one just by looking at them. Attribution of credibility by the source is a well documented practice in scientific literature. Due to the impossibility to fact-check every news article, researchers often relied on blacklists of low-credibility websites, curated by debunking organisations, in order to classify disinformation contents~\cite{Vargo2018TheAP,Allcott2019,guess2018selective,Tacchini2017}. Lazer et al.~\cite{Lazer1094} state that they ``advocate focusing on the original sources — the publishers — rather than individual stories, because we view the defining element of fake news to be the intent and processes of the publisher''. In fact, a malicious intentionality of the publisher is a distinctive characteristic of disinformation activities~\cite{CouncilofEurope}. Perception of the source can depend also on personal biases: Carr et al.~\cite{Carr2014} showed that sceptic and cynic users tended to attribute to citizen journalism website more credibility than to mainstream news websites. Go et al.~\cite{GOEun2014358} also pointed out that the credibility of an article is influenced by the believed expertise of the source, which in turn is affected by previous psychological biases of the reader, like in-group identity and bandwagon effect. In fact, also other people's judgement has been shown to produce an observable effect on individual perception of a news item~\cite{Houston2011, Rojas2010}. Lee et al.~\cite{Lee2010NeedForCognition} showed that people's judgement can be influenced by other readers’ reactions to news, and that others’ comments significantly affected one's personal opinion. Colliander et al.~\cite{COLLIANDER2019202} found that posts with negative comments from other people discourage users to share them. As noted by Winter et al.~\cite{Winter2015}, while negative comments can make an article appear less convincing to the readers, the same does not apply for a high or low number of likes alone. For an up-to-date review of the existing literature on `fake news' related research problems, see~\cite{ruffo2021surveying}.

\noindent
\textbf{Our contribution:} In our experiment we aimed to capture how the reader's decision process on the credibility of online news is modified by some of the above mentioned contextual features, which one was more effective on readers' mind, and on whom. We launched \texttt{Fakenewslab} (available at \\http://www.fakenewslab.it), an artificial online interactive news outlet designed to test different scenarios of credibility evaluation. We presented a sequence of 20 articles to volunteers, showing the articles to each user in five distinct ways. Some users find a title and a short description only, to simulate the textual information that would appear in the preview of the same article on social media. Additional information is presented to other users: some could read the full text of the article, someone else sees the source (i.e., a reference to the media outlet of the news), some other is told the percentage of other users that classified the news as true. Also, to better assess the significance of this last information, we showed a randomly generated percentage to a fifth group of users. The view mode is selected randomly for each volunteer when they enter the platform and start the survey. Everyone was asked to mark every single news as true or false, and this feature guided us in the preliminary selection of the articles, filtering out all those news where such a rigid distinction could not have been made.

Similarly, several other works asked online users to rate true or false news~\cite{Pennycook2019, micallef2021fakey, pennycook2018prior}, but to the best of our knowledge with the aim of quantifying how much a small twist in the interface of an online post can impact on people's evaluation of a news article. Our platform is inspired by MusicLab by Salganik et al.~\cite{salganik2006experimental}, an artificial cultural market that was used to study success unpredictability and inequalities dynamics, and that marked a milestone in assessing the impact of social influence on individual's choices. In particular, we adopted the idea of dispatching users in parallel virtual Rooms, where it is possible to control the kind and the presentation of the information they could read, in order to compare different behavioural patterns that may show high variability even for identical articles' presentation modes.  

Our results can have implications on the way we design social media interfaces, on how we should conduct news annotation tasks intended to train machine learning models whose aim is to detect ``fake news'', also on forthcoming debunking strategies. Interestingly, false news are correctly identified more frequently than true news, but showing the full article instead of just the title, surprisingly, does not increase general accuracy. Also, displaying the original source of the news may contribute to mislead the user in some cases, while social influence can positively assist individuals' ability to classify correctly, even if the wisdom of the crowd can be a slippery slope. Better performances were observed for those participants that autonomously opened an additional browser tab while compiling the survey, suggesting that searching the Web for debunking actually helps. Finally, users that declare themselves as young adults are also those who open a new tab more often, suggesting more familiarity with the Web.

\section{Data and Methodology}
\label{sec:methodology}
\texttt{Fakenewslab} is presented as a test accessible via Web, where volunteers are exposed to 20 news articles that have been actually published online in Italian news outlets; for each news, users should tell which were true or fake. Every user reads the same 20 news in a random order, and they are randomly redirected to one out of five different experimental environments, that we call ``virtual rooms'', in which the news are presented quite differently:

\begin{itemize}
    \item \textbf{Room 1} shows, for every news, the bare headline and a short excerpt, as they would appear on social media platforms, but devoid of any contextual clue.
    \item \textbf{Room 2} shows also the full text of the articles, as they were presented on the original website, again without any clear references to the source.
    \item \textbf{Room 3} shows the headline with a short excerpt and the source of the article, like it would appear on social media but devoid of social features. The article source is clickable and the article can be read on its original source.
    \item \textbf{Room 4} shows the headline with a short excerpt, and also the percentages of users that classified the news as true or false. We will refer to this information as ``social ratings'' from now on.
    \item \textbf{Room 5} is similar to Room 4, but with randomly generated percentages. We will refer to these ratings as ``random ratings'' from now on.
\end{itemize}

\begin{figure}[!ht]%
    \begin{subfigure}[t!]{0.5\linewidth}  
    \centering
    \includegraphics[width = \linewidth]{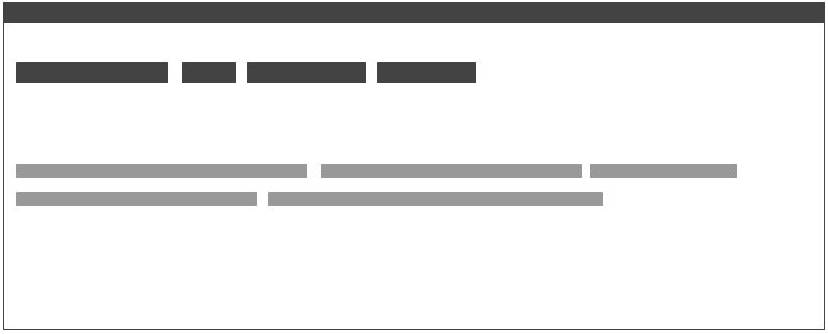} %
    \caption[]{Room 1: News headline and short excerpt.\newline}
    \end{subfigure}
    % \vskip\baselineskip
    \begin{subfigure}[t!]{0.5\linewidth}
    \centering
    \includegraphics[width = \linewidth]{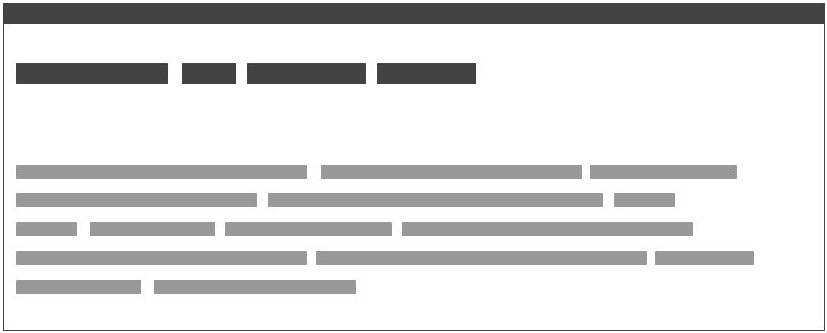}
    \caption[]{Room 2: News headline and full text.\newline}
    \end{subfigure}
    
    \begin{subfigure}[t!]{0.5\linewidth}  
    \centering
    \includegraphics[width = \linewidth]{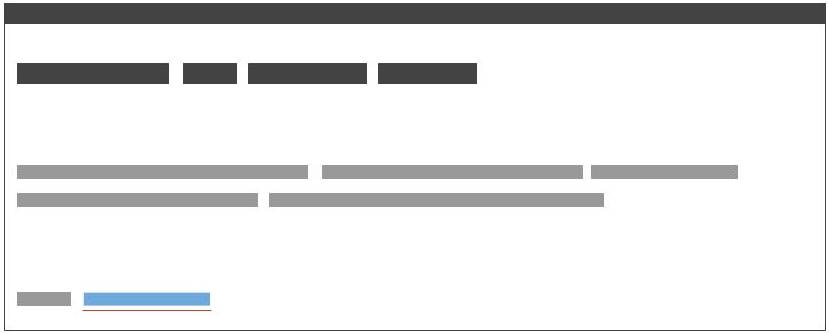} %
    \caption[]{Room 3: News headline and a clickable link\\ to the source.\newline}
    \end{subfigure}
    % \vskip\baselineskip
    \begin{subfigure}[t!]{0.5\linewidth}  
    \centering
    \includegraphics[width = \linewidth]{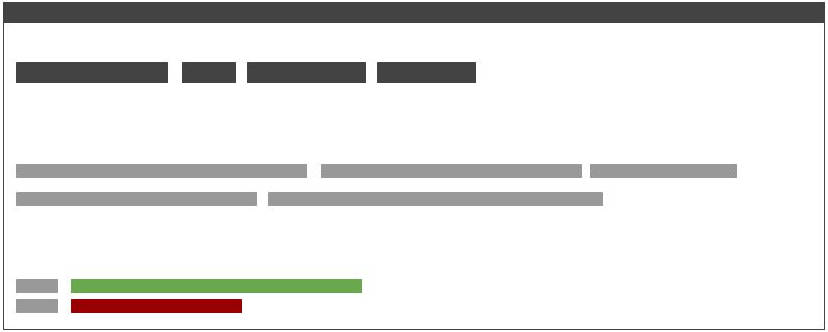} %
    \caption[]{Room 4: News headline and other users' ratings.\newline}
    \end{subfigure}
    
    \begin{subfigure}[b]{0.5\linewidth}  
    \centering
    \includegraphics[width = \linewidth]{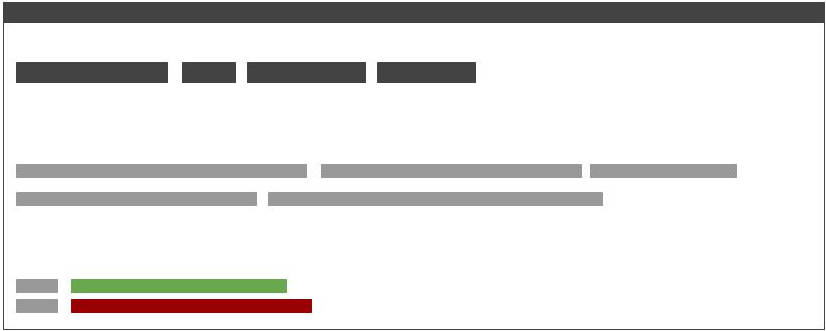} %
    \caption[]{Room 5: News headline and other users' ratings (random values).\newline}
    \end{subfigure}
    
    \caption{Graphical representation of the information displayed for each news in each Room.}
    \label{fig:rooms_aesthetic}%
\end{figure}

All the rooms are just variations of Room 1, that is designed as an empirical baseline for comparative purposes. Room 5, also, has been introduced as a ``null model'' to evaluate the social influence effect in comparison with Room 4, that displays actual values. Fig.~\ref{fig:rooms_aesthetic} displays a graphical render of how a news looked like in each Room.

During the test, we monitored some activities, in order to have a better insight on the news evaluation process. We collected data such as the timestamp of each answer (which allowed us to reconstruct how much time each answer took), what room the user was redirected to, and what social ratings or random ratings were they seeing in Room 4 and Room 5. We also measured if users left the test tab of the browser to open another tab during the test, an activity we interpreted as a signal of attempted fact checking: searching online in order to assess the veracity of an article was not explicitly encouraged nor forbidden by the test rules, and we left to the users the chance to behave like they would have done on everyday web browsing. We are aware that the mindset of a test respondent may have discouraged many users from fact checking the articles, as it may have been perceived like cheating the test, but our argument is that the very same mindset lead the users to pay an attention to news they would not have payed otherwise, thus raising an alert on possible deception hiding in each news: explicitly allowing for web searches out of the test would have probably lead to a disproportionate, unrealistic fact checking activity. We noted that eventually 18.3\% of users fact checked the news at least once on their own initiative, a percentage in accordance to existing literature~\cite{Hassan2017, Robertson2020}. However, estimating how many users would have fact checked the articles is outside the goals of this experiment: we only looked at those users who possibly did it, and at how fact checking impacted on their perception and rating of the news. Along with information related to the test, we asked participants to fill an optional questionnaire about demographics personal information. We collected users genre, age, job, education, political alignment, time and number of daily newspapers read, each of them on an optional basis. 64\% of users responded to at least one of the previous questions. This personal information helped us to link many relevant patterns in news judgement with demographic segments of the population. Details about such relationships between test scores and some demographic information are discussed in Appendix~\ref{app:demography}. 
All the users' data was anonymous: we did neither ask nor collected family and given names, email addresses, and any other information that could be used to identify the person that participated to the task. We used session cookies to keep out returning users to try the test more than once. However, the participants to the test were informed about the kind of data we were collecting and why (See Appendix~\ref{app:ethics}).
The test has been taken more than 10,000 times, but not everyone completed the task of classifying the 20 articles; in fact, although users were redirected to each room with a uniform probability, we observed ex-post that some them abandoned the test before the end, especially in Room 2 and partially in Room 5, as reported in Table~\ref{table:completion}. The higher abandon rate in these two rooms could be explained by the length of articles in Room 2, which may have burden the reader more than the short excerpts in other settings, and by the unreliable variety of ``other users ratings'' in Room 5 that were actually random percentages, that may have induced some users to question the test integrity. It should also be noticed that users that were redirected to Room 1 finished the test with the highest completion rate.

\begin{table}[h!]
\begin{center}
 \begin{tabular}{| c | c | c |} 
 \hline
 Room & Finished tests & Completion \\ [0.5ex] 
 \hline\hline
 \multicolumn{1}{|l|}{1: Headline} & 21.72\% & 81.76\% \\ 
 \hline
 \multicolumn{1}{|l|}{2: Full text} & 17.42\% & 70.00\% \\
 \hline
 \multicolumn{1}{|l|}{3: Source} & 20.81\% & 79.18\% \\
 \hline
 \multicolumn{1}{|l|}{4: Social ratings} & 20.50\% & 79.97\% \\
 \hline
 \multicolumn{1}{|l|}{5: Random ratings} & 19.55\% & 79.01\% \\ [1ex] 
 \hline
\end{tabular}
\caption{Tests completion percentages grouped by Rooms. Users are distributed randomly following an uniform distribution in the five Rooms, but complete tests in Room 2 are only 17.42\% of the total. Users in Room 2 left the test unfinished almost 10\% more times than users in other Rooms.}
\label{table:completion}
\end{center}
\end{table}

After some necessary data cleaning (i.e., leaving out aborted sessions and second attempts from the same user), we ended up with 7,298 unique participants that completed the 20 questions test (145,960 answers total). The answers were collected within a two weeks period in June 2020 by Italian speaking users. Fakenewslab was advertised by ourselves, through our personal pages on social networks such as Facebook and Twitter; it was subsequently advertised via word of mouth by the respondents that took the test. 

The test was followed by an ex-post Delphi survey, carried out by 10 selected experts of media, that responded to our answers about the main features that can have a role in the perception of credibility of news online. We designed the questions in order to provide a qualitative support to the main findings of our quantitative analysis. Results will be discussed in Section~\ref{sec:results}, while the methodological details and the list of questions is available in Section~\ref{app:delphi}. 

\subsection{News selection}
\label{sub:news_selection}
As mentioned in Sec.~\ref{sec:relatedworks}, the information disorders spectrum is wide, and it includes mis-/dis-/malinformation practices, that means that the reduction of the general problem to a ``true or false'' game is way too simplistic. 
In fact, professional fact checkers debunk a wide variety of malicious pseudo-journalism activities, flagging news articles in many different ways: reported news include not only those entirely fabricated, but also those that cherry-picked verifiable information to nudge a malicious narrative, or those that omitted or twisted an important detail, thus telling a story substantially different from reality. 

For our experimental setting, we selected 20 articles that we were able to divide into two equally sized groups of true (identified from now on with a number from 1 to 10) and false news (from 11 to 20) - see Appendix~\ref{app:news} for the complete list and additional information. Such division was straightforward in the majority of cases: we classified news as true when they were not falsified in any professional debunking sites and we also confirmed their veracity on our own. To challenge the user a little bit, we selected less known facts, sometimes exploited from some outlets as click-baiting, and sometimes with a poor writing style, with the exception of news 9 and 10, that we used as control questions. 
Marking a news as false can be much more ambiguous. However, we selected 10 stories that were dismissed in at least one debunking sites. In particular, article n. 14 was particularly difficult to classify as false for the reasons explained in App.~\ref{app:news}, but we kept it anyhow: in fact, when evaluating accuracy scores of \texttt{Fakenewslab} users, we are not calling out the gullibility of some of them, but we are only monitoring their activities and reactions that may be dependent on the environmental setting they were subject to. Thus it is important to stress here once again that we are more interested to the observed differences between Rooms' outcomes, than assessing people ability to guess the veracity of a news. Hence, even if in the following sections we will make use of terms such as ``correct'', ``wrong'', ``false/true positives'', or ``false/true negatives'' for the sake of simplicity, we must recall that we are measuring the alignment of users' decisions with our own classification just to assess those differences w.r.t. some given baseline.

% Results and Discussion can be combined.
\section{Results}
\label{sec:results}

\begin{figure}[h!]
    \centering
    \includegraphics[width = 0.7\linewidth]{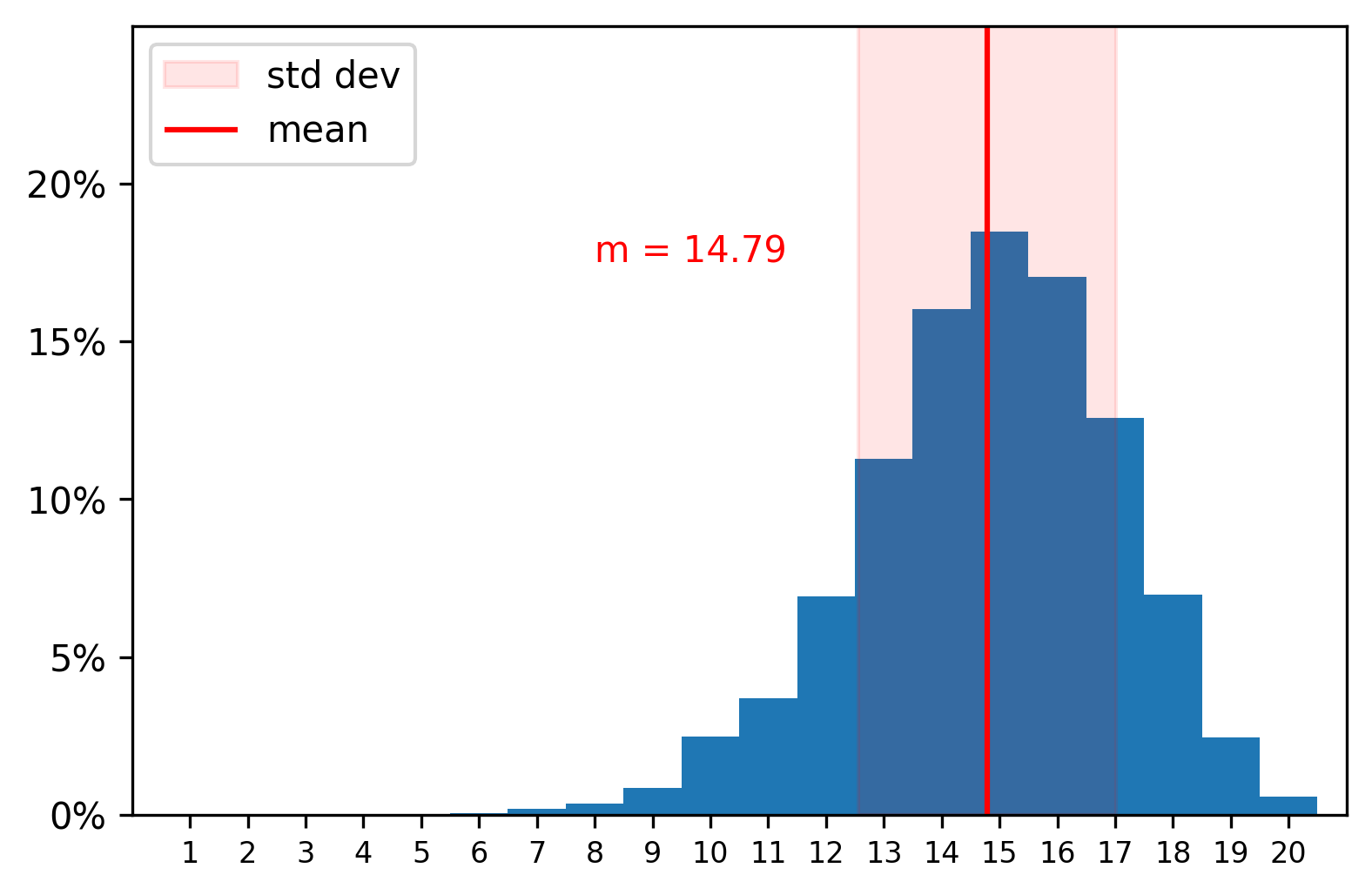} %
    \caption[]{Scores' distribution regardless the test settings. Users marked news in agreement with our own classification on average 14.79 times out of 20 questions. The most common score was 15/20.}
    \label{fig:overall_scores}
\end{figure}

\begin{figure}[h!]
    \centering
    \includegraphics[width =     0.5\linewidth]{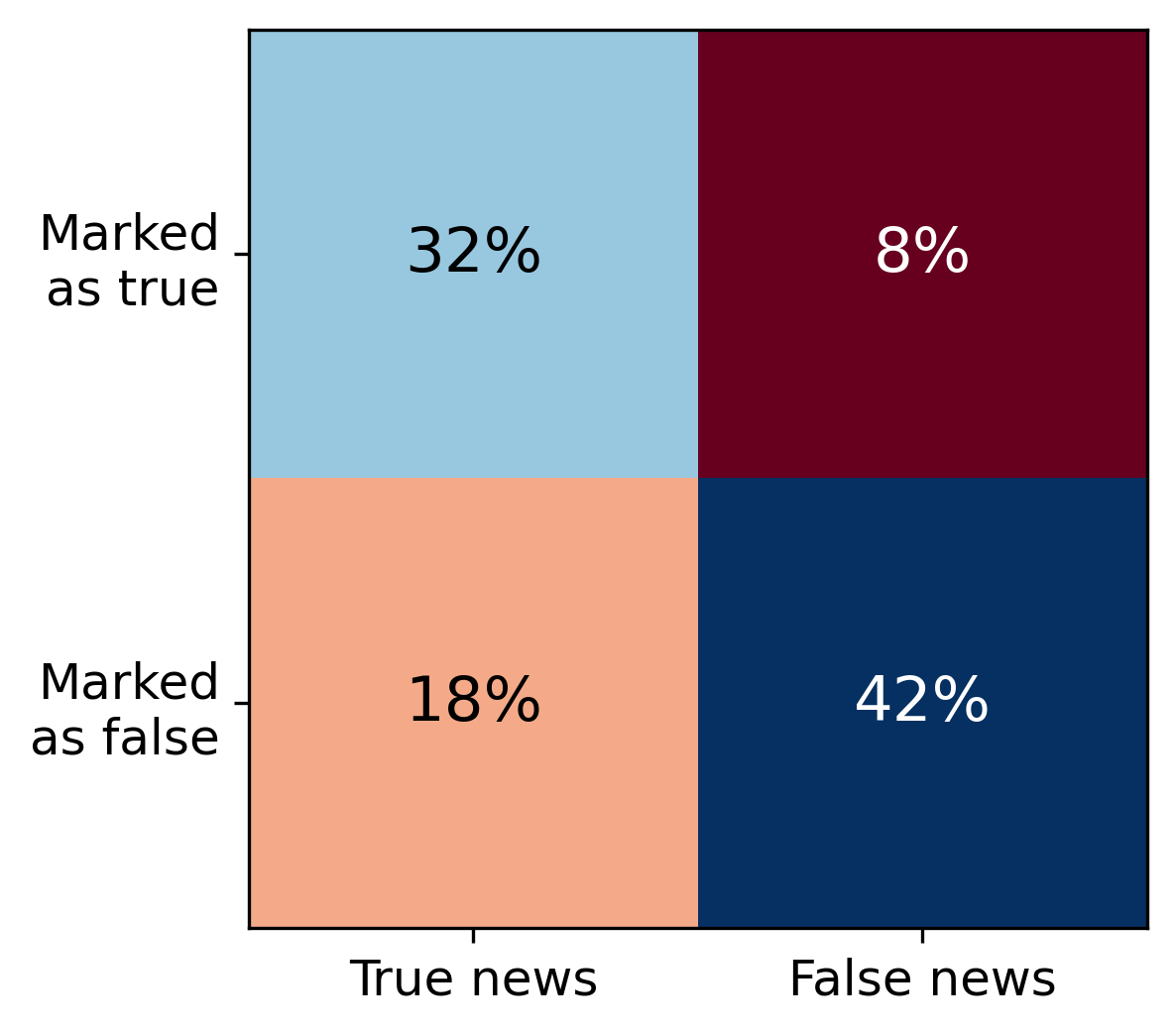}
    \caption[]{Users ratings' confusion matrix. Users rated a news as false 60\% of the times, i.e.,  marking as false even some true ones. True news have been often mistaken for false.}
    \label{fig:confmatrix}
    
\end{figure}

Overall, users that completed \texttt{Fakenewslab} showed a very suspicious attitude. 
Fig.~\ref{fig:overall_scores} and~\ref{fig:confmatrix} plot respectively the scores distribution and the confusion matrix of all accomplished tests, regardless of the setting (Room). Users marked news in agreement with our own classification on average 14.79 times out of 20 questions, with a standard deviation of 2.23. %

This quite generally ``good'' performance is unbalanced:
60\% of news were rated false, despite selected true and false news were actually equally distributed. False news, on the contrary, were hardly to be mistaken.
False negatives were therefore the most common kind of error (see Fig.~\ref{fig:confmatrix}). It looks like that users' high sensitivity during the test
lead to overall results similar to a random guessing game with a probability for a news to be false around three times out of four: in fact, a Mann-Whitney test does not reject the hypothesis that the overall distribution of correct decisions (i.e. accurate ratings of news) may be a binomial with probability $p \approx 0.74$.

\begin{figure}[h!]
    \centering
    \includegraphics[width = 1\textwidth]{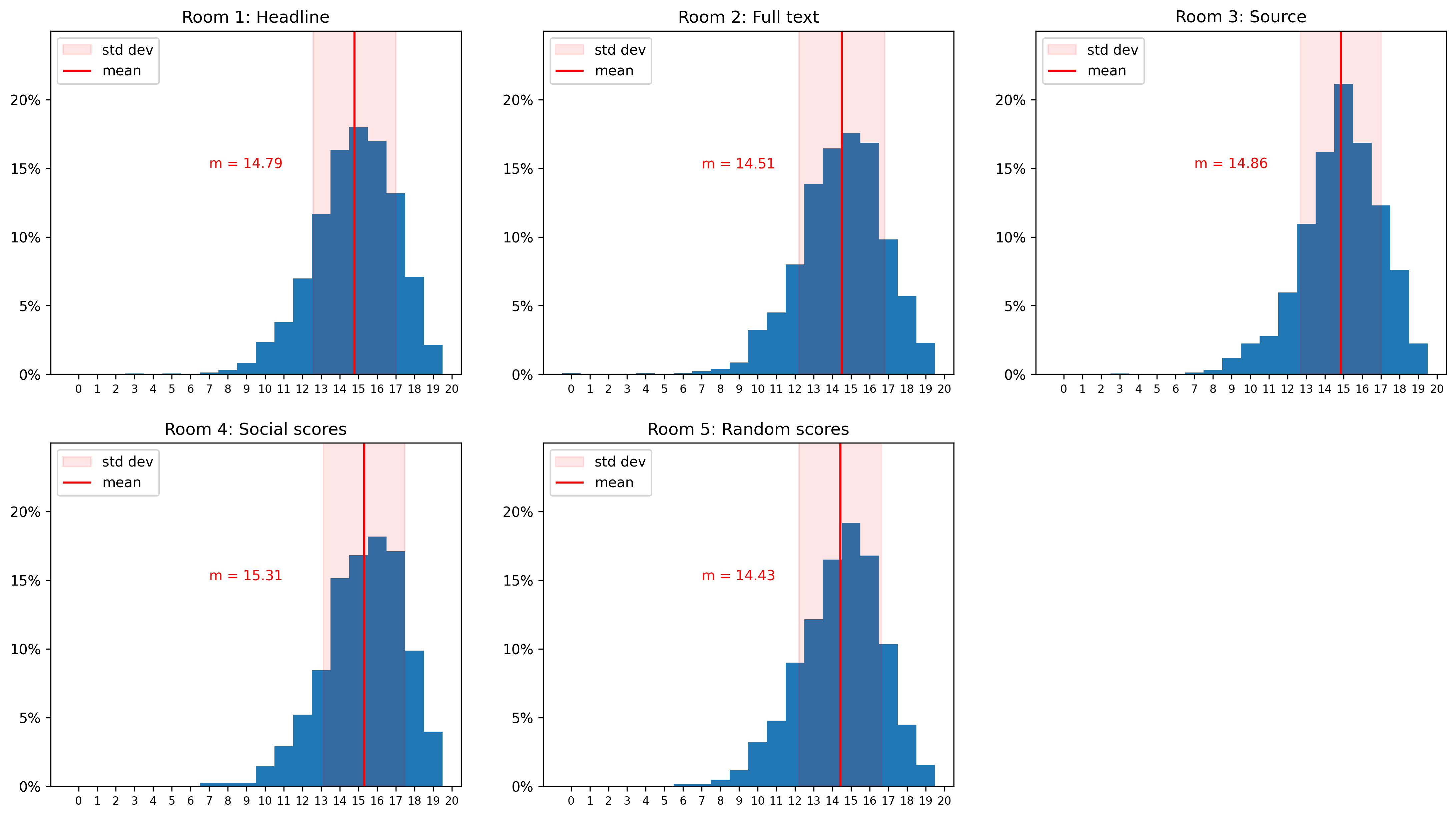}
    \caption{Scores' distribution by Room. Social influence is the feature with the highest positive/negative impact on users' scores: it can support the reader's judgement (Room 4), but it can also deceive it (Room 5).}
    \label{fig:results_by_room}
\end{figure}

Interestingly, there are differences depending on what kind of information the users were exposed to. Fig.~\ref{fig:results_by_room} shows the scores' distributions by Room. Users that saw the article's headline and its short excerpt (Room 1) scored an average of 14.79 correct attempts, matching exactly the overall average.

At first sight, all the other distributions and their characterising averages are similar to each other. Even so, if we keep Room 1 as a baseline, a Mann-Whitney test rejects the hypothesis that distributions for Rooms 2, 4, and 5 are equivalent to Room 1's (i.e., they are statistically significant diverse, with p-value $<10^{-5}$), while the same hypothesis for Room 3 could not be rejected. At a deeper insight, we observed that the article's source impacted the most on users' judgement; in fact, if we compute the average absolute difference between the users' decisions on each question between Room 3 and the other rooms, Room 3 is the one that produced the greatest gap. This effect was sometimes a positive, sometimes a negative contribution to users' decisions, depending on the specific news and source, making Room 1 and 3 scores distributions equivalent at the end. 

\begin{figure}[!ht]
    \centering
    \includegraphics[width = \linewidth]{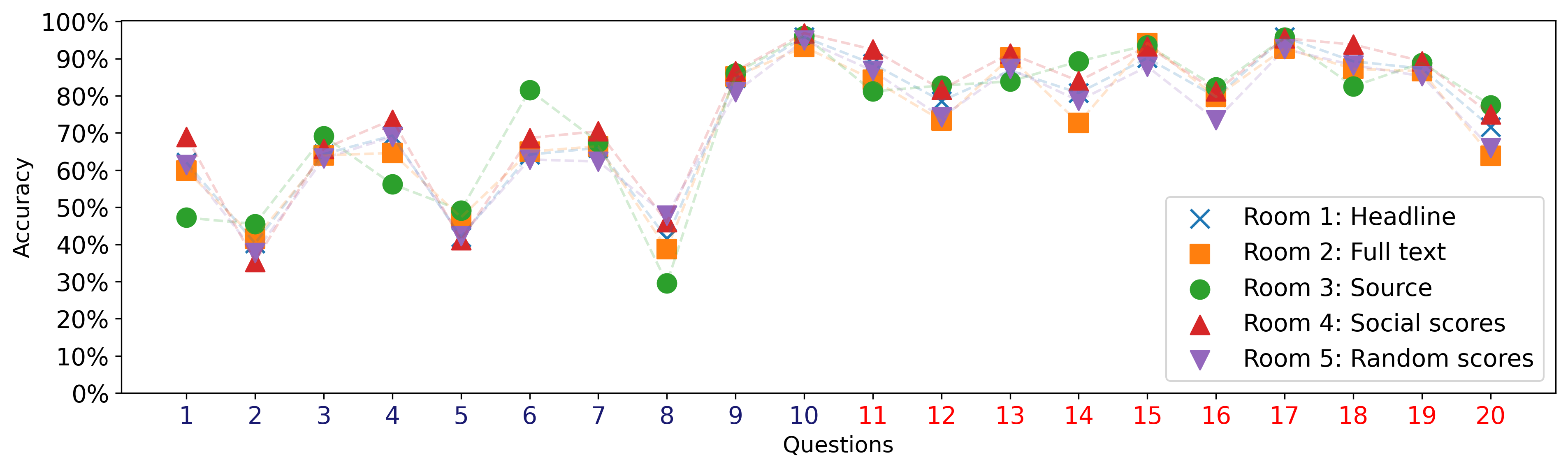}
    \caption{Accuracy for each question, grouped by Room. The source of the news (Room 3) is the feature that can influence users' judgement the most on the single question, either improving or worsening their scores, especially in questions 1, 4, 5, 8.}
    \label{fig:accuracy_question}
\end{figure}

This effect is shown in Fig.~\ref{fig:accuracy_question}, that displays how many times users made a correct decision on each news, grouped by Room. News are sorted here with true news first (blue x-axis labels) and false news after (red x-axis labels) for improving readability. While other users' ratings were the most helpful feature (Room 4, see Subsection~\ref{subsec:social_influence}), Room 3's users decisions are the ones that diverge the most from the others, especially for questions 1, 4, 6 and 8. Quite interestingly, articles 1, 4 and 8 reported true news, but the presentation had a click-bait tendency, taken moreover from online blogs and pseudo-newspapers (see Table~\ref{tab:newstags} in App.~\ref{app:news}), which probably induced many users to believe them false. Question 6, instead, reported a case of an aggression to policeman committed in a Roma camp, a narrative usually nudged by racist and chauvinist online pseudo-newspapers. The news was actually true, and the source mainstream (Il Messaggero is the eight most sold newspaper in Italy\footnote{Source: https://www.adsnotizie.it, accessed on April 29, 2021.}): users that read the source attributed to the news a very higher chance to be true. Rooms 2, 4 and 5 deserve a deeper discussion in the next Subsections.

\subsection{Individual decisions and social influence}
\label{subsec:social_influence}
 
The aim of Rooms 4 and 5 is to evaluate the impact of the so called ``Wisdom of the Crowd'' in users' decisions. Room 4 shows actual percentages of users marking every single news as true or false: here we observed the highest average score of 15.31. Room 5 has been created for validation purposes, since it displays percentages corresponding to alleged decisions by other users, that are actually random. This led to the observation of the lowest average score of 14.43 (see Fig.~\ref{fig:results_by_room}).

\begin{figure}[!ht]
    \centering
    \includegraphics[width = \linewidth]{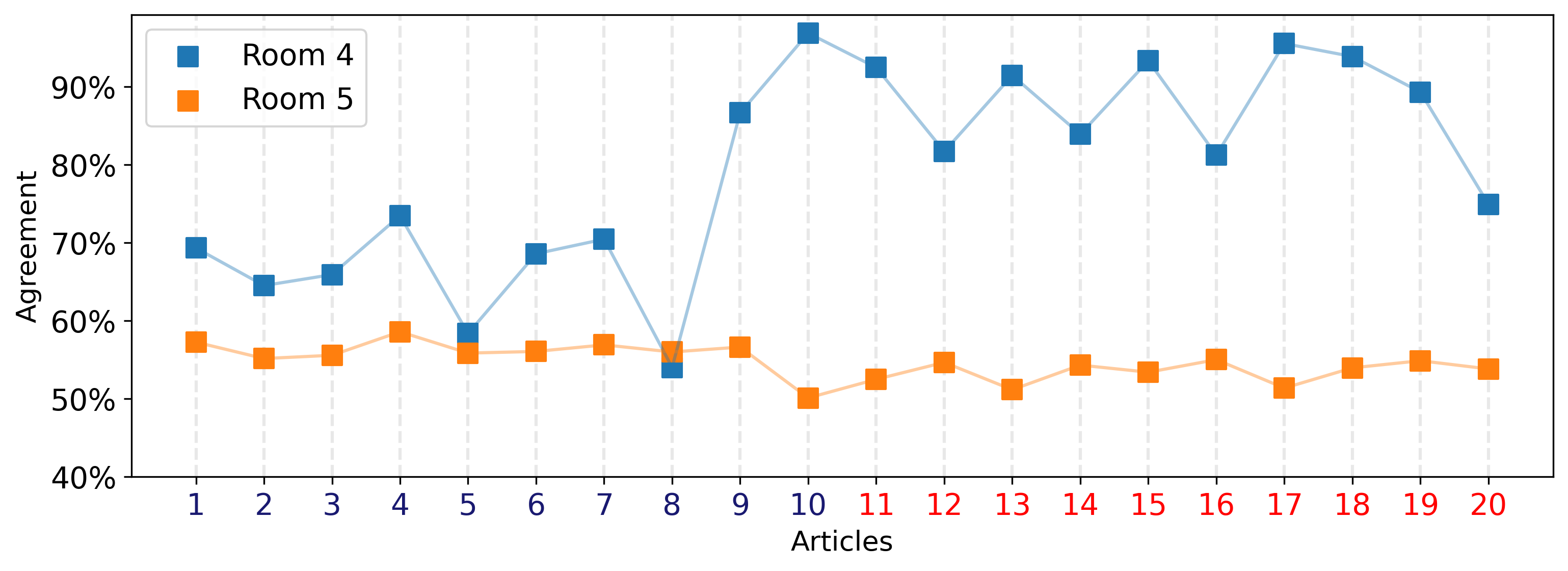}
    \caption{Agreement of users with the social ratings (Room 4, blue) and random ratings (Room 5, orange). Users conform to social ratings way more than random, inconsistent ones. Random ratings are likely trusted slightly more over true news, where users are probably more hesitating, than over false news. Similarly, social ratings were likely more helpful over false news than on true ones.}
    \label{fig:agreement}
\end{figure}

Our hypothesis on this important difference is that random ratings were sometimes highly inconsistent with the content of the news, maybe suggesting that an overwhelming quota of readers had voted true some blatantly false news, or the contrary. This inconsistency triggered suspects on the test integrity, which may be the primary cause for the slightly higher abandon rate we observed in Room 5 (see Table~\ref{table:completion}). When users did not abandon the test, they still were likely to keep a suspicious attitude. In line with our argument, Fig.~\ref{fig:agreement} shows the agreement of users with the ratings they saw for both Room 4 and Room 5. Overall, Room 4 users conform with the social ratings 79.28\% of times, while Room 5 users agreed with the random ratings only 54.66\% of times. This low value is still slightly higher than 50\%, i.e., a setting where users conform with the crowd randomly. As a result, there is a clear signal that their judgement was deceived as they performed worse than the users in other rooms. Users that saw the random ratings but did not follow the crowd performed exactly like users in Room 1, that is our baseline setting. 

It should also be observed that randomly generated percentages displayed with true news were taken more into account (55.8\% of the times) than random percentages shown with false news (53.5\% of the times), probably because we selected true news articles that were more deceptive due to their style and borderline content (see App.~\ref{app:news}). Room 5 users may have followed the crowd when insecure about the answer, and taken a decision independently otherwise, if the percentages contradict their own previous beliefs. On the other hand, users in Room 4 that saw real decisions from other users about true news may have followed the crowd much less than they did when deciding on false news (70.8\% against 87.7\%). Our hypothesis is that social ratings were perfectly consistent with strong personal opinions over false news, while the signal from other users over true news was a little bit noisy to their judgement. 

It is worth noting that the majority of respondents answered wrongly to questions 2, 5 and 8 (see Fig.~\ref{fig:accuracy_question}), thus probably posing to other users a dilemma: whether they should follow the crowd that was consistent and helpful other times, even if they disagree on the decision of the majority. For questions 5 and 8 the agreement with the social ratings are similar to the ones in Room 5, with random ratings (see Fig.~\ref{fig:agreement}), while the majority's behaviour on question 2 was accepted 64.5\% of times, even if related decisions were wrong.

\begin{figure}[!ht]%
    \begin{subfigure}[b]{\linewidth}  
    \centering
    \includegraphics[width =     \linewidth]{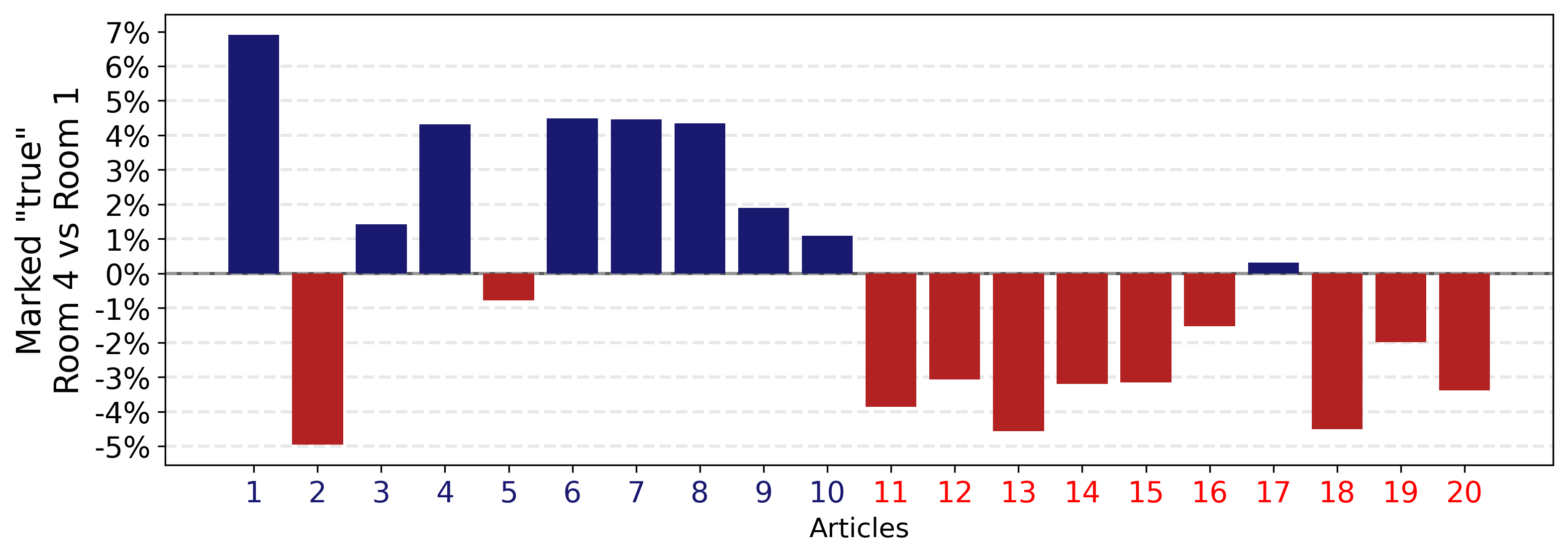} %
    \caption[]{Social ratings in Room 4 likely lead users to correct decisions, compared to those who did not have any additional information to use (Room 1).}
    \end{subfigure}
    \vskip\baselineskip
    \begin{subfigure}[b]{\linewidth}
    \centering
    \includegraphics[width =     \linewidth]{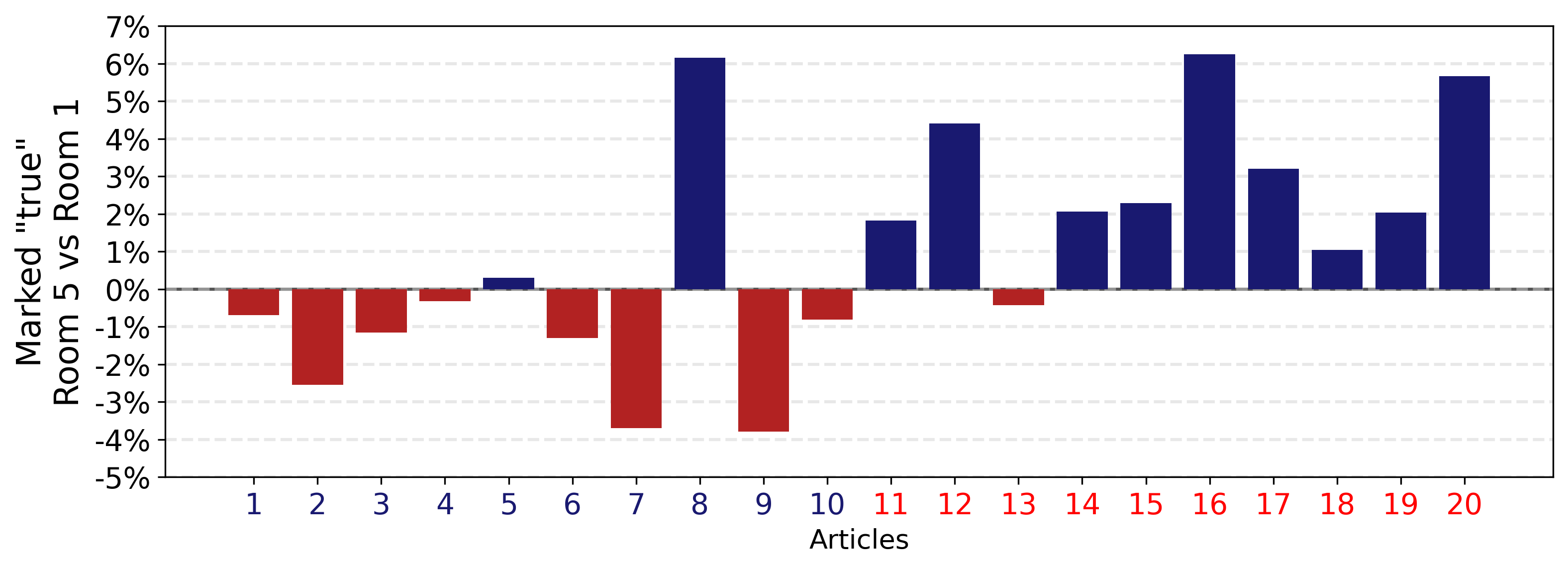}
    \caption[]{Random ratings in Room 5 probably affected users' judgement in a negative way, often deceiving them compared to those in Room 1.}
    \end{subfigure}
    \caption{Comparison of percentages of users that labelled news as ``true'' in Room 1 vs Room 4 (top) and Room 1 vs Room 5 (bottom).}
    \label{fig:normative_influence}%
    
\end{figure}

Fig.~\ref{fig:normative_influence} shows the percentage of users that rated news as true in Room 1 vs Room 4 (top)) and in Room 1 vs Room 5 (bottom). Users with no additional information tended to rate correctly true news fewer times, and rated false news as true more times, than users in Room 4. On the contrary, users in Room 5 were led to take true news as false, and otherwise. Social ratings and random ratings had a strong impact on individuals' judgement. This impact may be explained by a combination of the Bandwagon's effect~\cite{nadeau1993new, MyersBandwagon}, which suggests that the rate of individual adoption of a belief increases proportionally to the number of individuals that adopted such belief yet; and the social bias known as normative social influence~\cite{cialdini2004social, aronson2005social, COLLIANDER2019202}, that suggests that users tend to conform to the collective behaviour, seeking for social acceptance, even when they intimately disagree. \texttt{Fakenewslab}'s users, when redirected in Rooms 4 and 5, could have interpreted the social ratings and random ratings as a genuine form of wisdom of the crowd~\cite{Kremer2014}, and conformed to them, while users in Room 1 just answered according to their personal judgement.

\subsection{More information is not better information}
 \label{subsec:full_text}
 
Articles' headlines and short excerpts are designed to catch the readers' attention, but full details of the story can only be found delving deep in the body of the article. The body of an article is supposed to report details and elements that bring information to the story: location, persons involved, dates, related events, commentaries. However, when the users could read the full text of the news in Room 2 instead of merely a short headline, their judgement was not helped by more details, but instead hindered: average scores in Room 2 are lower than the overall case, with 14.51 correct decisions per user (see Fig.~\ref{fig:results_by_room}). Fatigue could be an explanation. Table~\ref{table:completion} reports lower completion percentages for users in Room 2, suggesting that many of them could have been annoyed while reading longer stories, thus abandoning the test prematurely. Also, users that completed the test usually took less time to answer, if compared with the other Rooms: the median answer time is about one second less than the overall median answering time. 
 
\begin{figure}[!ht]%
    \begin{subfigure}[b]{\linewidth}  
    \centering

    \vskip\baselineskip
    \begin{subfigure}[b]{\linewidth}
    \centering
    \includegraphics[width = 0.8    \linewidth]{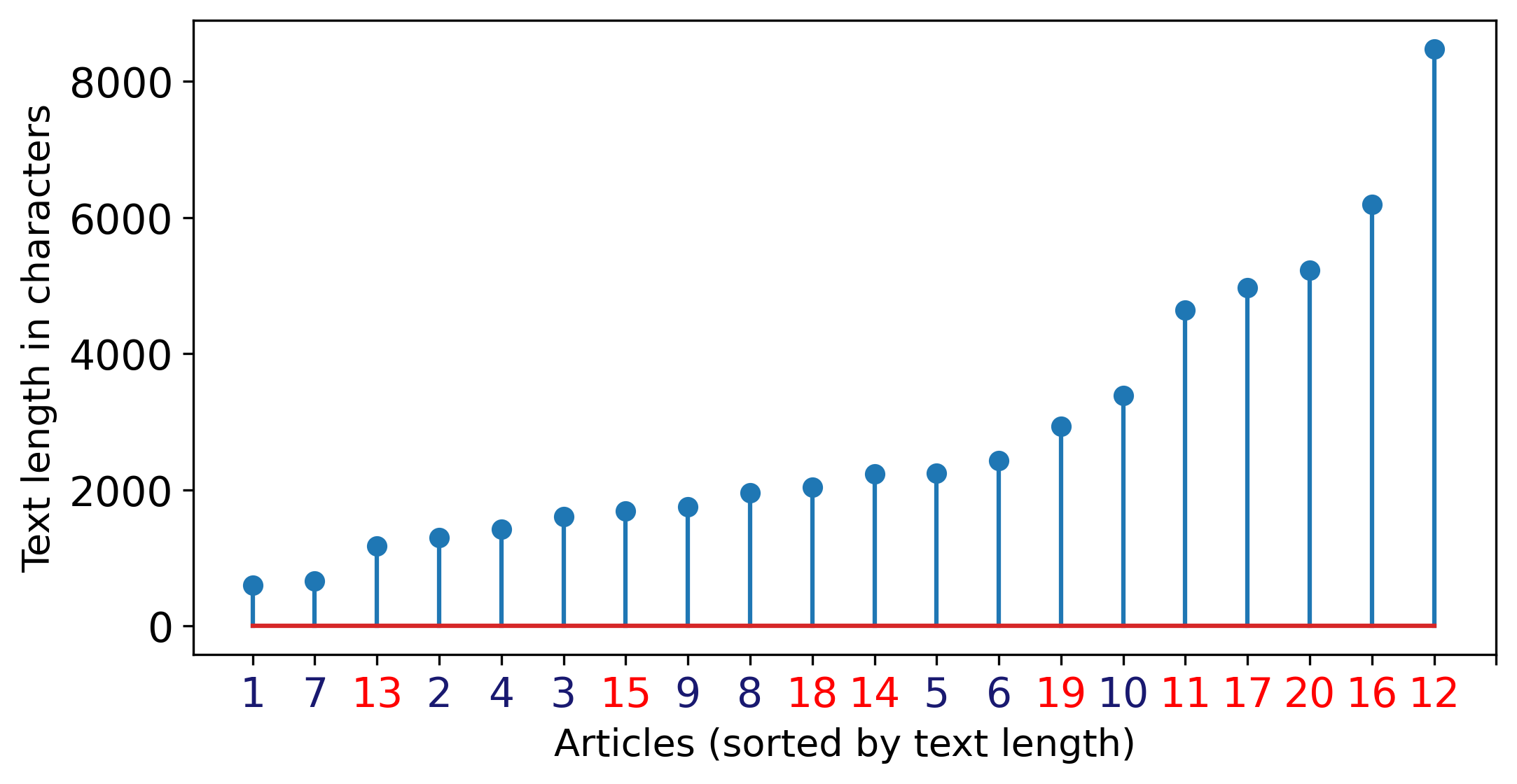}
    \caption[]{Length of texts in Room 2. Articles' bodies are sorted from the shortest to the longest.}%
    \end{subfigure}
    \includegraphics[width = 0.8    \linewidth]{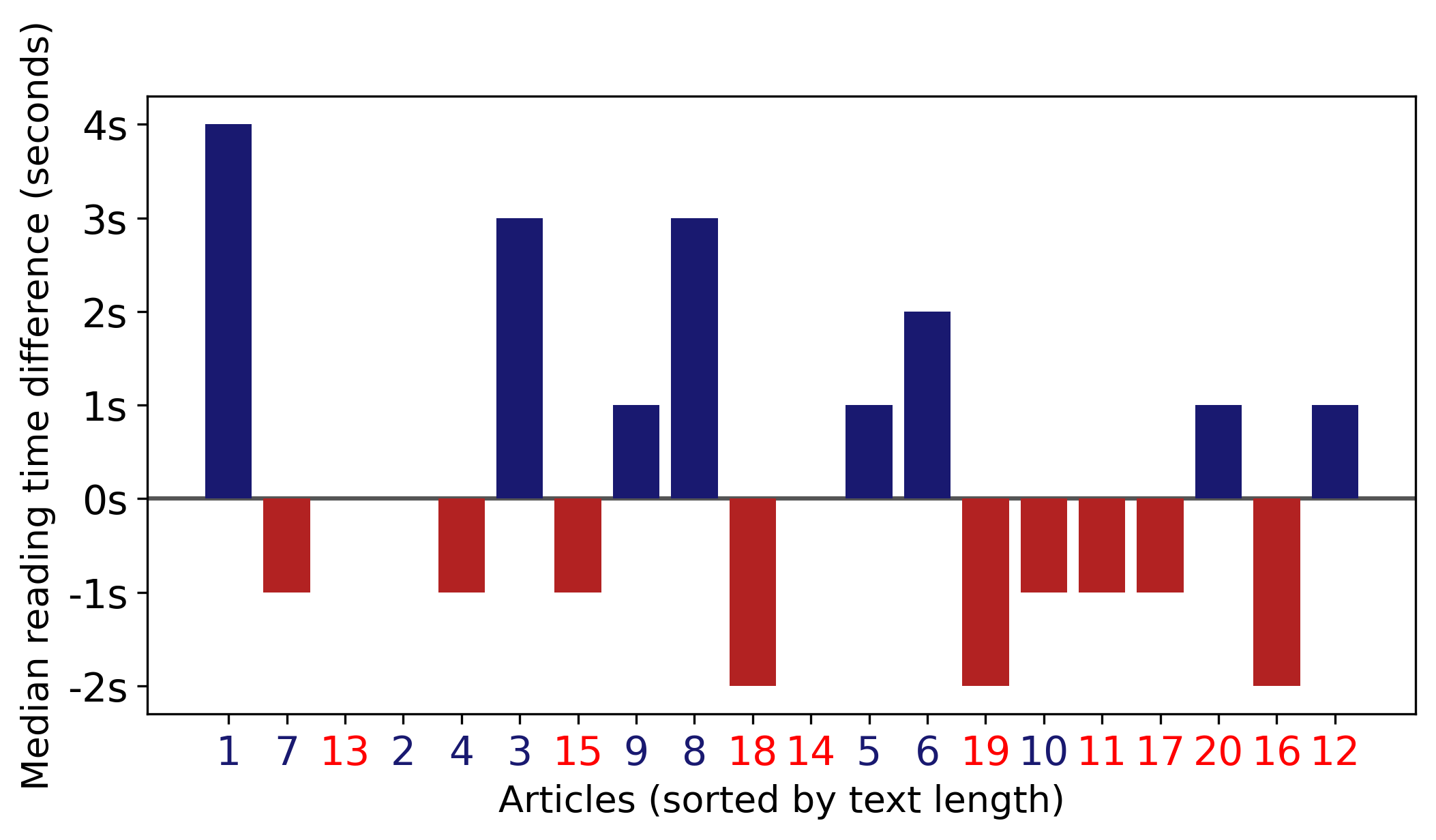} %
    \caption[]{Difference between the median answering time taken by users in Room 2 and all the other rooms on each article, with articles sorted by length. 
    Apparently, there is no linear dependency of answering times on text length. Also, the decision on the longest articles shown in Room 2 took less time in 5 cases out of 7.}
    \end{subfigure}
    \caption{Articles full body length are not correlated with longer answering times.}
    \label{fig:text_length}%
    
\end{figure}

In fact, reading time does not depend on the length of the text, as shown in Fig.~\ref{fig:text_length}: on the top we show the articles length in characters, that ranges from few hundreds to more than 8,000. Longer articles are way longer than shorter ones, and they would require much more time to be read entirely. However, this was not observed. In the bar-plot on the bottom of Fig.~\ref{fig:text_length}, we plot the difference between the median answering time in Room 2 and in all the rooms altogether, grouped by question. The bar is blue when the difference is positive, i.e., users in Room 2 took more time to answer than the other users, while it is red otherwise. Questions in Fig.~\ref{fig:text_length} are sorted from the shortest to the longest. Answering time does not appear related to the text length, but rather to the news itself. However, from \nth{14} position and beyond, bars are more likely red (negative difference): users decided more quickly with longer texts when they had the opportunity to read them, suggesting that they did not read the texts entirely, possibly because they made their mind in the headline anyway, or also because the writing style of the full article just confirmed their preliminary guesses. 

\begin{figure}[!ht]%
    \begin{subfigure}[b]{\linewidth}  
    \centering
    \includegraphics[width =  0.8   \linewidth]{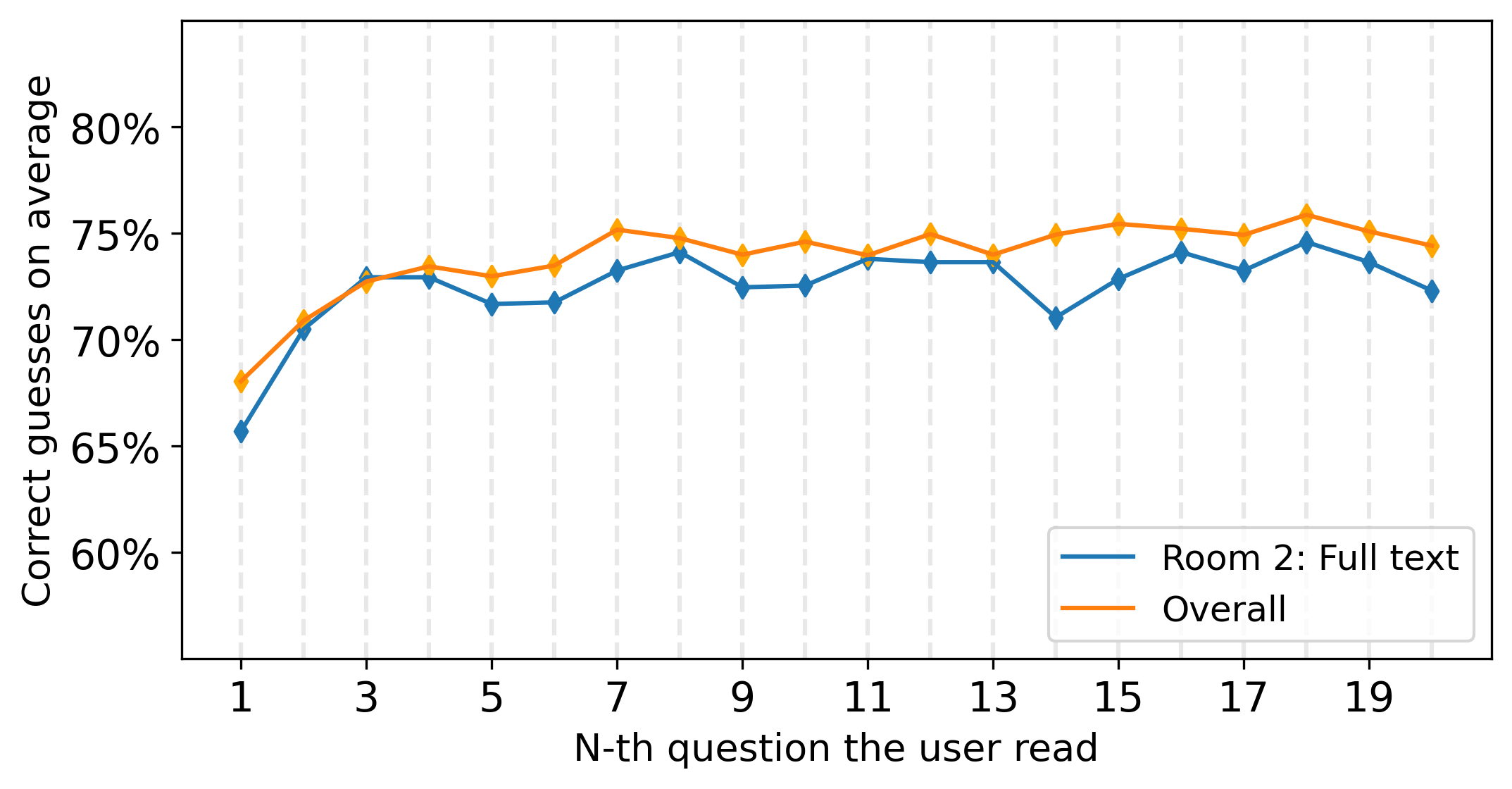} %
    \caption[]{Accuracy in Room 2 does not drop progressively. However, it is constantly lower than overall accuracy.}
    \end{subfigure}
    \vskip\baselineskip
    \begin{subfigure}[b]{\linewidth}
    \centering
    \includegraphics[width = 0.8     \linewidth]{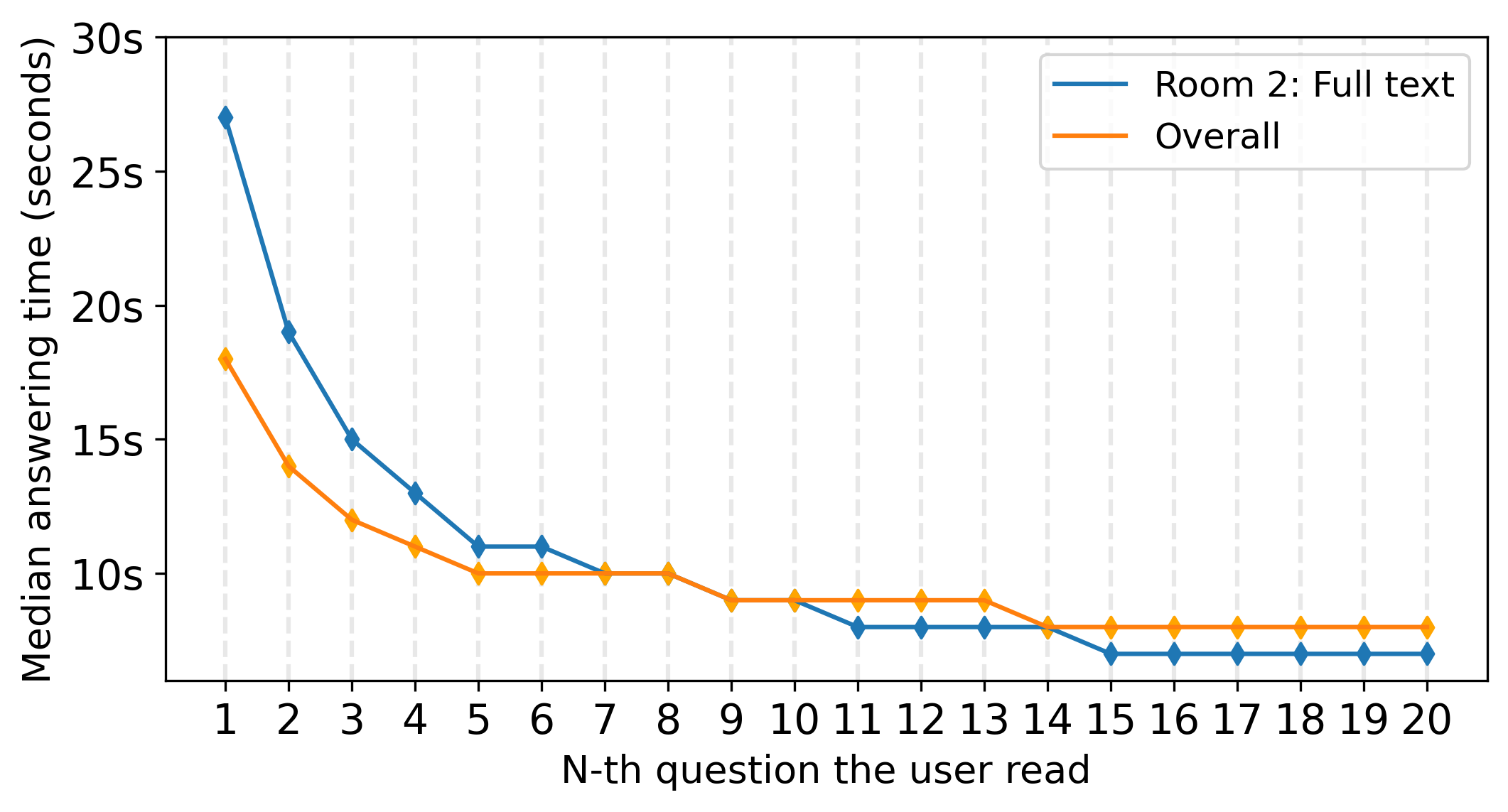}
    \caption[]{Users in Room 2 start reading the full text on the first questions, but they get quickly annoyed and the reading time fall in short. During the second half of the test, Room 2 users answer even more rapidly than in the other rooms.}%
    \end{subfigure}
    \caption{Average accuracy (top) and median answering times (bottom) of users for questions in order of presentation to the users. Values in Room 2 are compared to their corresponding values in all the rooms taken altogether.}
    \label{fig:fatigue}%
\end{figure}

Moreover, users in Room 2 tended to answer more quickly to the last half of the test, if compared to users that only saw a shorter text. In Fig.~\ref{fig:fatigue}, we reported the median answering time in both Room 2 and overall, with questions ordered as the articles were actually presented to the reader. On the first questions of the test, users in Room 2 took more time to answer, apparently reading the long texts. However, their attention collapsed quickly, and from the \nth{7} question and beyond they spent even less time reading the articles than the users in other Rooms. The relationship between the users' fatigue and the average scores is not clear, however. The right panel shows the average accuracy of users by question, where questions are ordered according to the reader's perspective. If a cumulative fatigue was burdening the users of Room 2 more than the others, average accuracy in the last questions would have to progressively fall. In the top panel of Fig.~\ref{fig:fatigue}, this effect is not evident, even if a slight degrading of accuracy is visible. The interplay between fatigue, texts length and user's behaviour must be properly addressed in a dedicated experiment, but these preliminary findings suggest that the toll required by a long read in terms of attention may discourage users from actually reading and processing more informative texts, and that may lead to even wronger decisions. The length of an article has been shown to be an informative feature for artificial classifiers that discriminate between true and false news~\cite{Kumar2016,Ghanem2020}; our results suggest that this can not be the case for human evaluators. As a consequence, the design of any AI whose purpose is to automatically classify news should take into accounts such limitation.

\subsection{Web's familiarity, fact checking and young users}
 \label{subsec:opened_tabs}
 
One common way to assess the veracity of a piece of news is online fact-checking, i.e., searching for external sources to confirm or disprove the news. In Room 3 the users could see the sources of the articles, and a link to the original article was displayed: we observed that 26.5\% of the users clicked at least once on such a link during the test. However, fact-checking activity requires the user to check news against external sources; retrieving the original web article could only confirm its existence, and give some clue about the publisher (for instance, by reading other news from the same news outlet, or by technological and stylistic features of the website~\cite{Chung2012}). True fact-checking would have required the user leaving temporarily the test, to browse in search for confirmation on other websites~\cite{Wineburg2017LateralRR}. As anticipated in Sec.~\ref{sec:methodology}, we captured information about the users that left the test for a while, coming back at it a few seconds later, by monitoring when the test tab in the users' browser was no longer the active one. It is worth recalling that we had no way to see what the users were searching in other tabs, if they were actively fact-checking the test articles or just taking a break from the test. Also, in our guidelines we did not explicitly mention the possibility of searching the answers online.

However, estimating how many users do fact-checking of online news is outside of the scope of this work, and it is a research problem that must be addressed properly. We only acknowledge that opening a new tab and coming back to the test later could be a hint of an external research; that we did nor encouraged nor discouraged such activity; and that in the end 18.25\% of the users did open another tab at least once during the test. 
 
  \begin{figure}[!ht]
    \centering
    \includegraphics[width =0.7\linewidth]{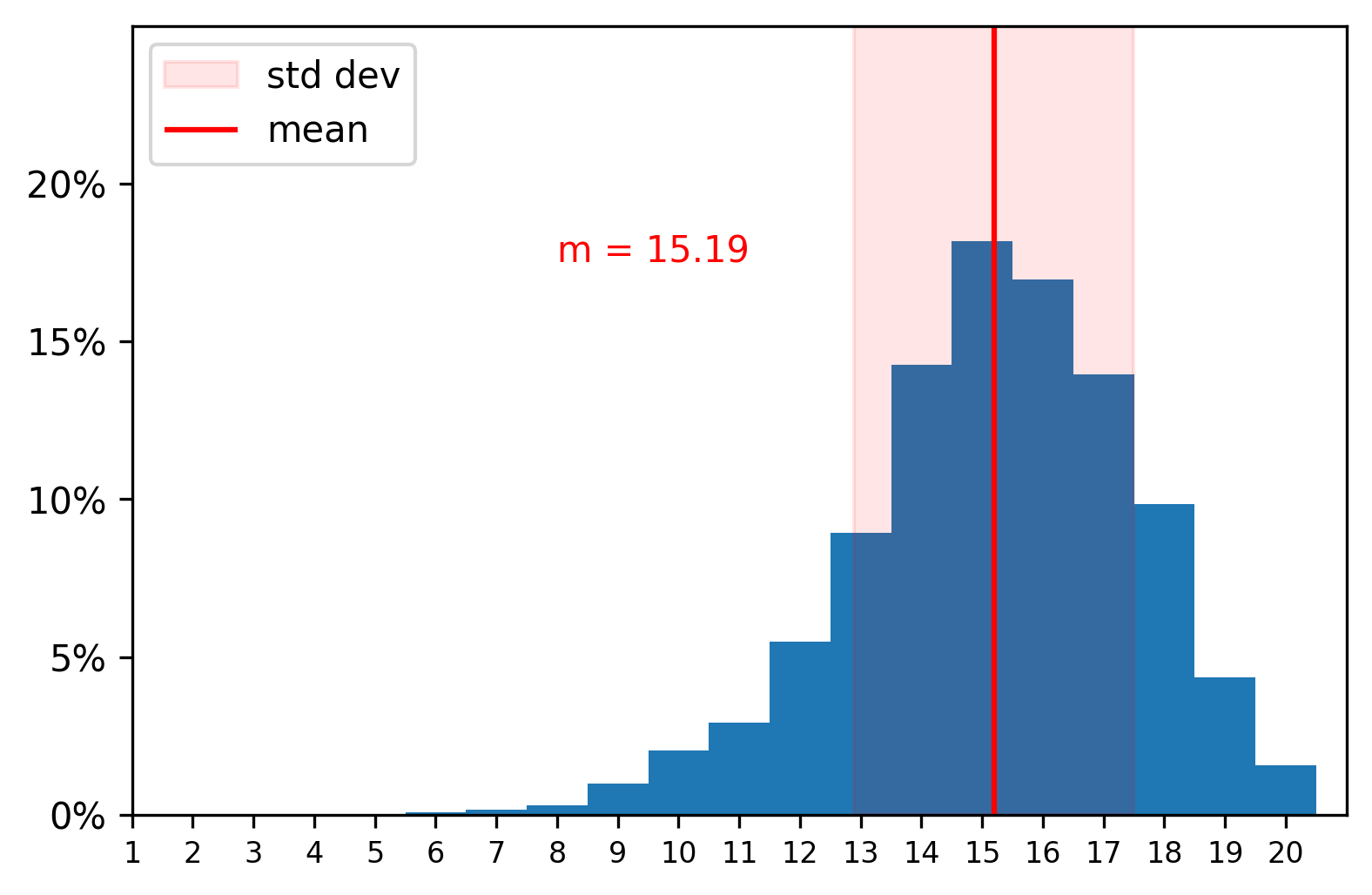}
    \caption{Distribution of scores for those users that opened another tab during the test, and returned to the test a while after, at least once. Actually, the score's average is higher here than the overall outcomes (see Fig.~\ref{fig:tab_user_scores}). Tab opening could be a hint of online fact-checking.}
    \label{fig:tab_user_scores}
\end{figure}

 \begin{figure}[!ht]
    \centering
    \includegraphics[width=0.7\linewidth]{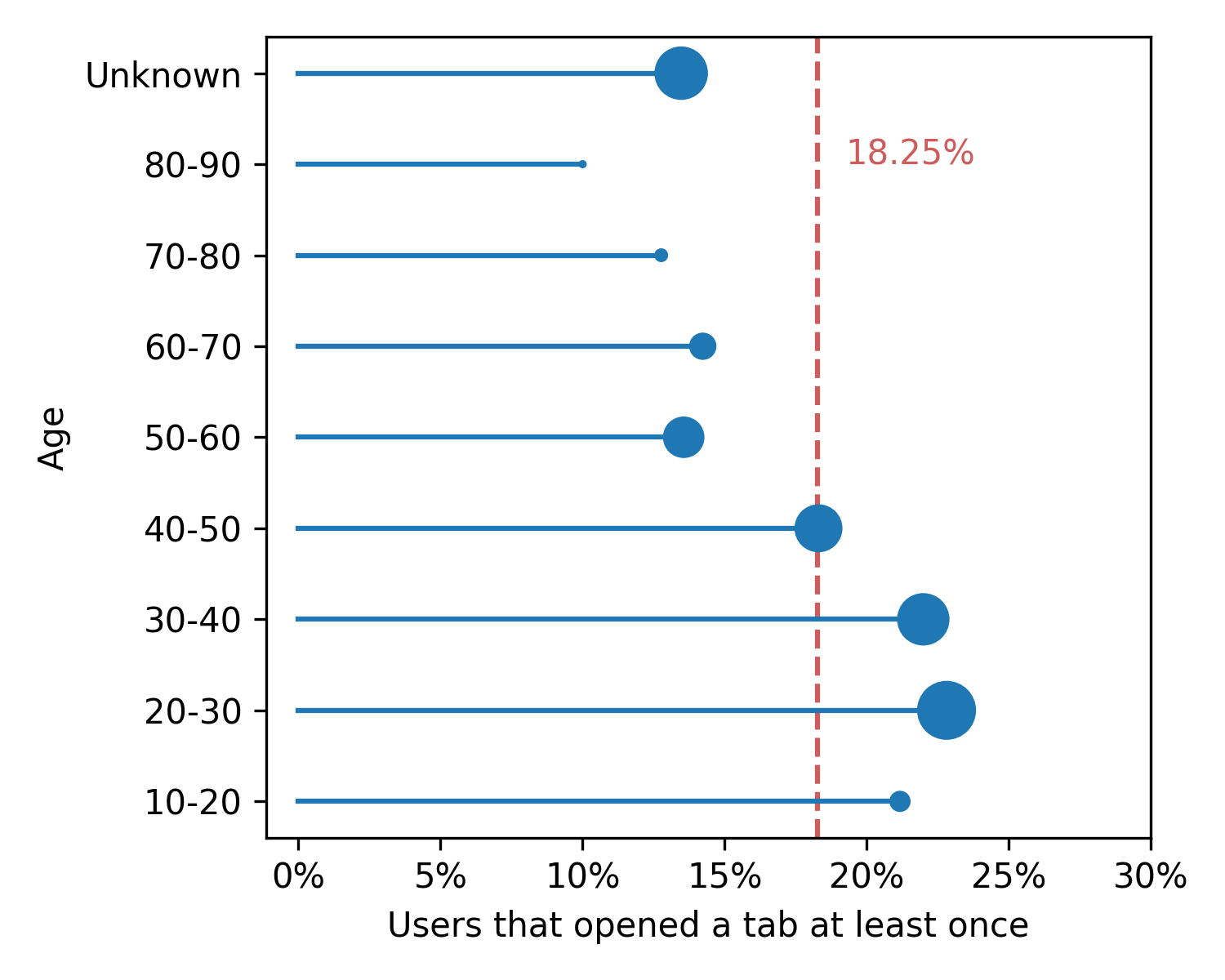}
    \caption{Percentage of users that opened a tab at least once, grouped by age. Lollipop size is proportional to the subgroup population in our dataset. On average, 18.25\% of users opened another tab at least once. Teenagers and young adults in the groups 10-20, 20-30, 30-40 opened a tab more often than the others.}
    \label{fig:who_tabs}
\end{figure}

Users who opened a tab answered correctly 78.23\% of times, against the 73.82\% of the others that did not. Fig.~\ref{fig:tab_user_scores} reports the average distribution of users' scores, only for those who have opened a tab at least once. The mean of the distribution (15.19 correct guesses over 20) is close to the one in Room 4, significantly higher than in Room 1. Opening a tab for searching an answer online is an activity that we therefore correlate with a possible fact-checking, due to the positive effects it has on users' tests. Also, it is an activity that correlates with the age of the respondents. Fig.~\ref{fig:who_tabs} shows the percentage of users that opened another tab in the browser, by age. Each age category is represented by the corresponding lollipop, which is long proportionally to the tab opening rate, and sized after the number of respondents in such age category. The vertical dashed red bar points out the percentage of users who opened a tab, regardless of their age. Young adults and adults in the age 20-30 through 50-60 are the most represented in the data, but their tab opening rate is different. Users in the age groups 20-30 and 30-40 showed a higher propensity to open another browser's tab, while users in the 50-60 age group did it significantly less than the average. Other categories are under-represented, but they still confirm the general trend: overall, the chances for users to open a new tab while responding to the test are higher for teenagers and young adults, and they decrease substantially with age. Incidentally, young adults and teenagers are the users that are often considered showing more familiarity with technologies such as web browsers~\cite{Brashier2020}, the ones that more easily could think of opening a parallel research in another tab. Wineburg et al.~\cite{Wineburg2017LateralRR} noted that the most experienced users about online media, when called to tell reliable online news outlets from unreliable ones in test conditions, tended to open a new tab. This result suggests that users with higher familiarity with the medium are more prone to fact-check dubious information online, carrying a simultaneous and independent research on the topic on their own, even when not encouraged to do it. Young adults are also the age group with the highest average scores. 

% \section{Discussion}
% Nulla mi mi, venenatis sed ipsum varius, Table~\ref{table1} volutpat euismod diam. Proin rutrum vel massa non gravida. Quisque tempor sem et dignissim rutrum. Lorem ipsum dolor sit amet, consectetur adipiscing elit. Morbi at justo vitae nulla elementum commodo eu id massa. In vitae diam ac augue semper tincidunt eu ut eros. Fusce fringilla erat porttitor lectus cursus, vel sagittis arcu lobortis. Aliquam in enim semper, aliquam massa id, cursus neque. Praesent faucibus semper libero~\cite{bib3}.

\section{Conclusion}

\label{sec:conclusion}

Via \texttt{Fakenewslab} volunteers were asked to mark as true or false 20 news, previously selected from some online news outlets. We monitored users' activities under five different test environments, each showing articles with distinct information. Users could see the news piece in one of the following forms: the bare headline with a short excerpt of the article; the headline and the full article; the headline with a short excerpt and the original source the article was taken from; how other users rated the article (social scores); a randomly generated rating, presented like how other users rated the article (random scores). We also asked 10 experts of media to qualitatively answer 10 ex-post questions that would help us validating our findings.

The attribution of credibility to news based on the perceived credibility of its source is a well-documented practice in literature~\cite{GOEun2014358,Carr2014,bhagat2022examining}. This is especially true for political content produced by partisan news outlets, which are perceived to be more or less credible depending on whether they are ideologically congruent with prior political beliefs~\cite{turel2021biased,robertson2021trust}. Experts that responded to our Delphi survey (see Section~\ref{app:delphi}) also rated the source of an article as a highly informative feature. However, participants able to see the news's source did not classify articles better than those who have only seen the headlines, rather the source misled many users, probably because users stopped focusing on the news itself, distracted by the (good or bad) reputation of the outlet. In case of borderline or divisive news, such as in our test, the reputation of the venue may have obfuscated the inherent quality of the news article.

Although longer texts could be more informative than shorter ones, our findings suggest that reading an entire article instead of a short preview does not help in distinguishing false from true news. Common sense would suggest that in the full text the reader could find hints and details to better support their decision; this misconception was also confirmed by the panellists we consulted in our Delphi survey (see Section~\ref{app:delphi}) and by experimental results~\cite{wobbrock2021goldilocks}. However, users that could read the full text performed even worse. A possible explanation for this phenomenon, supported by our analyses, may lie in the fatigue of reading numerous long texts. 

Volunteers exposed to others' decisions were helped by the ``wisdom of the crowd'', though this could be a slippery slope: in fact, manipulated random feedbacks pushed a negative herding effect leading to more mis-classifications. Peer influence is acknowledged in literature as capable of impacting individual evaluations of credibility of news articles~\cite{bhagat2022examining}, especially in the form of bad comments~\cite{waddell2018does,petit2021can}. Interestingly, the panel of experts that responded to our survey was divided about the effectiveness of peer influence on one user's judgement (see Section~\ref{app:delphi}): 30\% of them rated other users' influence as non relevant to their own evaluation.

Last, young adults were prone to interrupt momentarily the test, coming back at it with more accurate answers, a possible hint of an underlying fact-checking activity. This is due, in our opinion, to higher media literacy and web tools expertise among young generations. Millennials and Generation Z users could be more encouraged to carry out a ``lateral reading'', i.e., searching the web for relevant information to validate news articles they come across, either by early exposure to the pitfalls of disinformation on internet or by their education~\cite{breakstone2021lateral,brodsky2021associations}. The relationship between users' young age and their propensity to lateral reading was also acknowledged by 60\% of the expert panellists that responded to the ex-post survey.
Quite interestingly, 30\% of them did not believe that age could be a discriminant in reading and fact-checking habits, while 20\% indicated Generation X (born between 1965 and 1980) as the generation of users more likely to fact-check a piece of news. Experts also underestimated the percentage of users that would carry out parallel research when reading news (see Section~\ref{app:delphi}).

The present study has some limitations that we care to point out. The test was advertised through social networks (Facebook, Twitter) personal pages of authors first, then in large public groups and via sponsored advertisement. As discussed in Appendix~\ref{app:demography}, respondents may show significant biases in the distribution of age, education, gender and political leaning, reflecting the demographics of our acquaintances instead of a neutral sample. Another limitation lies in the way we monitored users' activity. We broadly discussed in method the impossibility to check on users' activity when they left the test, to get back at it later. We did not encouraged nor discouraged users to fact-check the news, yet we speculate some of them searched some news on the web: we could verify that many of them left the test and got back at it later, and those who did it performed significantly better than others. 
% Last, we did not collect any information about the device they were using. Although we developed a mobile-friendly version of the desktop website, we did not control for a possible confounding given by the user experience of the mobile version versus the desktop version. 

Despite these caveats, the results here reported may support drawing up forthcoming guidelines for annotation tasks to train fake news classification systems, and for properly designing web and social media platforms. Such findings challenge the widespread source-based and crowd-based approaches for automatically distinguishing false from true news. Although low credibility online articles posted on social media is often on display as responsible for mis-/dis-information spreading, high reputation news outlets are accountable, too, because their inaccurate publications can have a greater impact on reader's minds. Furthermore, our findings indicate that suggesting users to read an article in its entirety, as currently done by AI-generated messages by social networks platforms, can backfire because of the heavy cognitive toll requested to readers. 

Further research is needed: e.g., the correlation between reading long texts and poor decisions has to be better addressed. Social influence is an important asset, but we are also aware that informational cascades are vulnerable to manipulations: 
studying the underlying mechanisms on how users evaluate news online, individually or influenced by other factors, is crucial in fighting information disorders online.

\section{Conflicts of Interest}
The authors declare no conflict of interest.

\newpage
\appendix

\section{Supplementary Materials}
\label{app:suppmaterials}

\subsection{Ethical and Reproducibility Statement}
\label{app:ethics}
All the collected data was shared by users on a voluntary basis, on the clear premise that it would have used for research purposes only. Users were not pressured in taking the test in any way. The motivation for the experiment was explained before the test. The existence of different test settings and the monitoring of the users activity, instead, were disclosed to respondents only after the test, as it would have severely affected the results. Users filled the questionnaire on their personal information on a voluntary basis: it was told them that they could have skipped the questionnaire and immediately access the results of the test. All collected information is anonymous, as we did not store personal information such as family or given names, email or IP addresses, or any other data that could be used to identify the person that participated to the task. We set in users' browsers a cookie, which they consented to, in order to detect a second attempt to the test that we ignored in our analyses. Last, it was clarified that data would never be shared with other subjects, and that it would have been used for research purposes only. All the collected information will be distributed \emph{in aggregated form} in a dedicated Github repository, for reproducibility purposes.

\subsection{Demography of users}
\label{app:demography}

\begin{figure}[!ht]
    \centering
    \includegraphics[width = \textwidth]{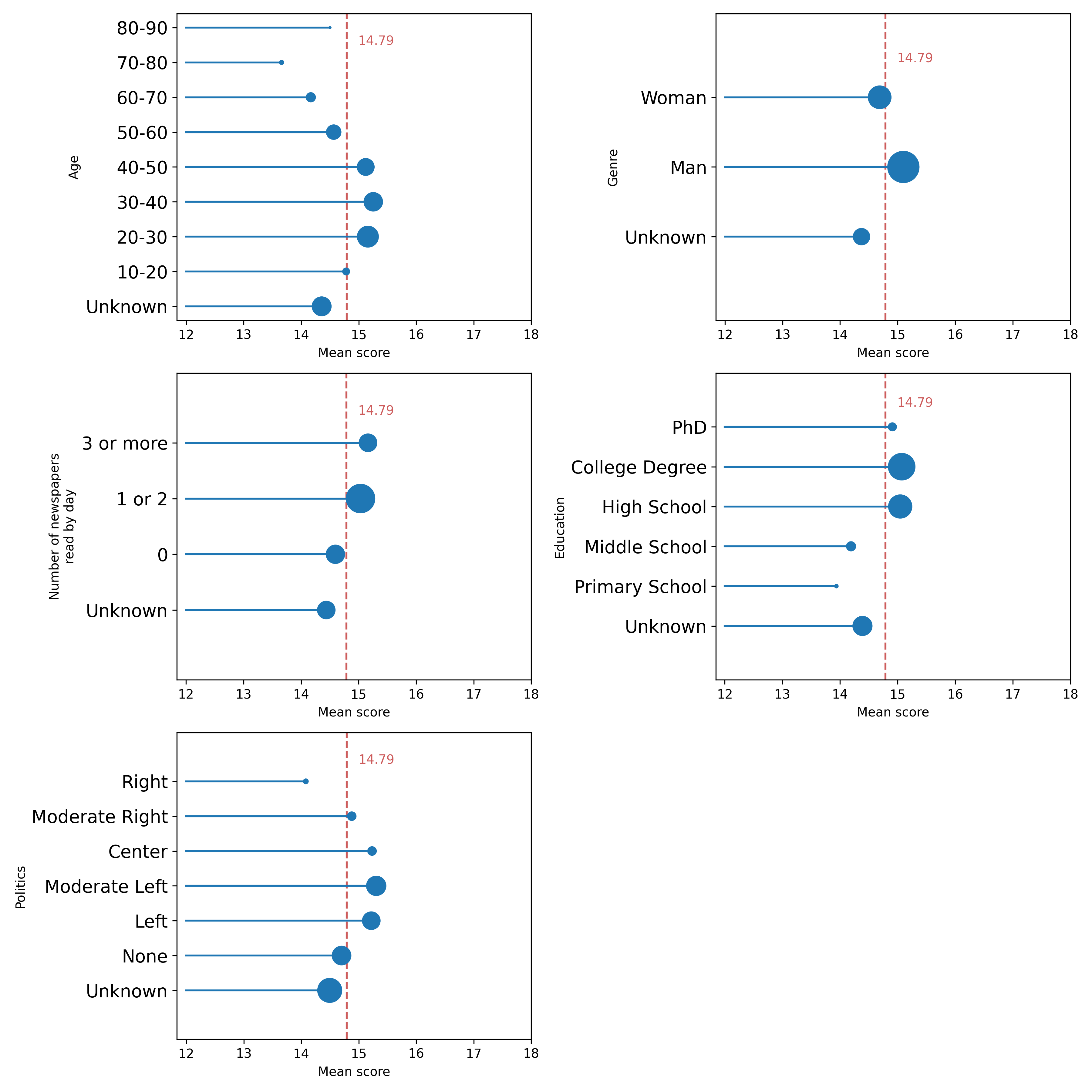}
    \caption{Scores' distribution by four personal information that users filled into the optional questionnaire. Each lollipop is sized proportionally to the population.}
    \label{fig:scores_anag}
\end{figure}

Along with their evaluation of news, we asked \texttt{Fakenewslab} users to respond to an optional questionnaire about their personal information. This questionnaire was taken by 64\% of users. In Fig.~\ref{fig:scores_anag} we report the average scores of users depending on some details they declared, like their age, genre, the number of newspapers they read by day, their education and their political leaning. Population segments are represented by lollipops. Each lollipop's length is proportional to the accuracy of users in that segment, while its size is proportional to their number. Due to the word of mouth spreading process of \texttt{Fakenewslab} on social networks, the population show some biases, as men are over-represented in comparison to women, and the same applies to college degrees (or even PhDs) over high school (and lower) diplomas, and to left and moderate left voters over moderate right and right. Even with this limitations, it is worth noting that mean scores are higher for young adults and adults from 20 to 50 years old; for users that declared to read at least three newspapers per day, over those who read 1, 2 or none; for users with higher school diplomas and degrees. Although scarcely represented, it is interesting that users that declared to have pursuit a PhD performed slightly worse than users that declared to have a college degree or a high school diploma. Conservative political attitude has been demonstrated as a factor correlated to believing and sharing disinformation content\cite{Grinberg374, Guess2019}. Right voters in this study performed an average score significantly lower than left voters, yet the heavily imbalanced distribution of the two sub-populations make it impossible to substantiate this claiming, as we could not properly stratify segments of users based on politics and genre or age.

\subsection{News}
\label{app:news}

We presented to \texttt{Fakenewslab} users 20 news, and asked them to decide whether the news were true or false. However, as described in Sec.~\ref{sub:news_selection}, the ``true or false'' framing is an oversimplification of a faceted problem. News flagged by fact checkers include click-baits, fabricated news, news with signs of cherry-picking, crucial omissions that frame a story into a new (malicious) narrative, and other subtle stretches of truth. On the other side, true news can also be borderline, open to a broader discussion. In order to offer a complex landscape of debatable information, we did not choose only plausible news from famously reliable sources, which would have been probably easier to spot by contrast with false news, but we also selected a few borderline news, true news taken from blogs, some with even a slight tendency to click-bait. The final sample included hence sometimes borderline and ambiguous pieces of information to process. In tables~\ref{tab:newslist} we listed the 20 news proposed in \texttt{Fakenewslab}, and in Table~\ref{tab:newstags} we added some information on the source the debunking.

    \begin{table*}[h!]
    \centering
    \begin{adjustbox}{width=\linewidth}
    \begin{tabular}{|c|c|l|}
    \hline
    & Id & Headline \\
        \hline \hline
    \parbox[t]{2mm}{\multirow{10}{*}{\rotatebox[origin=c]{90}{True news}}}
    &  1 & A ``too violent arrest'': policeman sentenced on appeal trial to compensate the criminal\\
       \cline{2-3}
    & 2 & The Islamic headscarf will be part of the Scottish police uniform\\
        \cline{2-3}
    & 3 & Savona, drunk and drug addict policeman runs over and kills an elderly man\\
        \cline{2-3}
    & 4 & Erdogan: The West is more concerned about gay rights than Syria\\
        \cline{2-3}
    & 5 & \# Cop20: the ancient Nazca lines damaged by Greenpeace activists\\
        \cline{2-3}
    & 6 & Rome, policemen attacked and stoned in the Roma camp\\
        \cline{2-3}
    & 7 & Thief tries to break into a house but the dog bites him: he asks for damages\\
        \cline{2-3}
    & 8 & These flowers killed my kitty - don't keep them indoors if you have them\\
       \cline{2-3}
    & 9 & Climate strike, Friday for future Italy launches fundraising\\
        \cline{2-3}
    & 10 & Paris, big fire devastates Notre-Dame: roof and spire collapsed | The firefighters: ``The structure is safe''\\
        \hline \hline
    \parbox[t]{2mm}{\multirow{10}{*}{\rotatebox[origin=c]{90}{False news}}}
    & 11 & Carola Rackete: ``The German government ordered me to bring migrants to Italy'' \\
        \cline{2-3}
    & 12 & With the agreement of Caen Gentiloni sells Italian waters (and oil) to France \\
        \cline{2-3}
    & 13 & Vinegar eliminated from school canteens because prohibited by the Koran\\
        \cline{2-3}
    & 14 & He kills an elderly Jewish woman at the cry of Allah Akbar: acquitted because he was drugged\\
       \cline{2-3}
    & 15 & The measles virus defeats cancer. But we persist in defeating the measles virus!\\
       \cline{2-3}
    & 16 & 193 million from the EU to free children from the stereotypes of father and mother\\
        \cline{2-3}
    & 17 & Astonishing: parliament passes the law to check our Facebook profiles\\
        \cline{2-3}
    & 18 & Italy. The first illegal immigrant mayor elected: ``This is how I will change Italian politics''\\
        \cline{2-3}
    & 19 & INPS: 60,000 IMMIGRANTS IN RETIREMENT WITHOUT HAVING EVER WORKED\\
        \cline{2-3}
    & 20 & EU: 700 million on 5G, but no risk controls\\
        \hline
    \end{tabular}
    \end{adjustbox}
    \caption{News headlines' list. In our experiment, we adopted 20 articles corresponding to 10 true and to 10 false news. They were actually published in Italian online outlets, mainstream or not. In all the rooms, we integrally reported headlines, short excerpts, but the full articles were available only in Room 2. In this table, we show the (translated) headlines.}
    \label{tab:newslist}
    \end{table*}
 
    \begin{table*}[h!]
    % \begin{adjustwidth}{-2.25in}{0in} 
    \centering
    \begin{adjustbox}{width=\linewidth}
    \begin{tabular}{|c|c|l|c|l|}
    \hline
    & Id & Source & Type & Tagged as \\
        \hline \hline
    \parbox[t]{2mm}{\multirow{10}{*}{\rotatebox[origin=c]{90}{True news}}}
    & 1 & Sostenitori delle Forze dell'Ordine & blacklisted & \\
        \cline{2-4}
    & 2 & Il Giornale & mainstream &  \\
        \cline{2-4}
    & 3 & Today & mainstream &  \\
        \cline{2-4}
    & 4 & L'antidiplomatico & online newspaper &  \\
        \cline{2-4}
    & 5 & Greenme & online newspaper &  \\
        \cline{2-4}
    & 6 & Il Messaggero & online newspaper &  \\
        \cline{2-4}
    & 7 & CorriereAdriatico & mainstream &  \\
        \cline{2-4}
    & 8 & PostVirale & blog &  \\
        \cline{2-4}
    & 9 & Adnkronos & mainstreeam &  \\
        \cline{2-4}
    & 10 & TgCom & mainstreeam &  \\
        \hline \hline
    \parbox[t]{2mm}{\multirow{10}{*}{\rotatebox[origin=c]{90}{False news}}}
    & 11 & IlGiornale & mainstreeam & Wrong Translation -- Pseudo-Journalism \\
        \cline{2-5}
    & 12  & Diario Del Web & online newspaper & Hoax -- Alarmism\\
        \cline{2-5}
    & 13 & ImolaOggi & blacklisted & Hoax \\
        \cline{2-5}
    & 14 & La Voce del Patriota & blog & Clarifications Needed \\
        \cline{2-5}
    & 15 & Il Sapere è Potere & blackisted & Disinformation \\
        \cline{2-5}
    & 16 & Jeda News & blacklisted & Well Poisoning  \\
        \cline{2-5}
    & 17 & Italiano Sveglia & blacklisted & Hoax -- Disinformation \\
        \cline{2-5}
    & 18 & Il Fatto Quotidaino & blacklisted & Hoax \\
        \cline{2-5}
    & 19 & VoxNews & blacklisted & Unsubstantiated -- Disinformation  \\
        \cline{2-5}
    & 20 & Oasi Sana & blog & Well Poisoning -- Pseudo-Journalism \\
        \hline
    \end{tabular}
    \end{adjustbox}
    \centering
    \caption{Articles' additional information. We show the sources for each article of Table~\ref{tab:newslist}; in fact, even if the same news could have been reported in more different news outlets, in Room 3 we showed only the source that published the article as it was presented to our volunteers. Moreover, we give here some more information, such as the type of the news outlet (i.e., mainstream, online newspaper, blog, or blacklisted by some debunking sites), and the tags used by fact-checkers upon correction. Such information was not explicitly disclosed to the users.}
    \label{tab:newstags}
    % \end{adjustwidth}
    \end{table*}
 
Every news shown in Table~\ref{tab:newslist} has been fact-checked in at least one outlet among debunking sites or mainstream newspaper from the following list: \texttt{bufale.net}, \texttt{butac.it}, \texttt{ilpost.it}, \texttt{open.online}. According tags adopted by fact-checkers, we classified news from 11 to 19 as false (see~Table~\ref{tab:newstags}). News n. 20 is somehow an exception: we found it on an online website focused on health, openly against the 5G technology. Apparently, the news did not became viral, and it has been ignored by mainstream journalism and also by debunkers. The article suggests that the European Union has not assessed the health risks associated to the 5G technology, which is clearly false~\footnote{e.g., see the ``Health impact on 5G'' review supported by the European Parliament \texttt{https://www.europarl.europa.eu/RegData/etudes/STUD/2021/690012/\-EPRS\_STU(2021)690012\_EN.pdf}.}, and full of misunderstanding, to say the least, on how the European Commission funding processes really work.
 
It should be noticed that the most controversial classification, among the ``false'' news, is probably article n. 14. The source, ``La Voce del Patriota'', is a nationalist blog. The article refers of a factual event: a drugged Muslim man killed a elderly Jewish woman, and the reported homicide is mainly framed as racial, serving a broader narrative that criminalise immigration. When it was published the trial was only at its first stage, and the acquittal was temporary, waiting for a rebuttal. However, a correction of the news was available at the time of our experiment in some debunking sites; in fact, further investigations dismissed the racial aggravating factors, stressing the drug-altered state of the murderer. This is an example of how difficult could be to answer to a simple ``true or false'' game. The fact was true, but it was reported incorrectly. Nevertheless, we should recall here that our objective is not to judge people's ability to tell true from false, but how a social media platform's contextual information may influence us from making such decision. Hence, the results we observed comparing user's activities in different rooms are the core of our motivations.

\subsection{Qualitative assessment of biases about perception of credibility}
\label{app:delphi}

Aiming to pair the quantitative analysis we performed in Section~\ref{sec:results} with qualitative feedback, we deployed an ex-post survey, asking 10 media experts to share their experience and their beliefs about how they perceive news articles online, depending on the auxiliary information they can read. The methodology we used is the Delphi method~\cite{Delphi}: it consists of several rounds of survey (typically two or three) where domain experts are called to answer questions anonymously, preventing that any of the experts could influence their peers due to their authority. Between two rounds, a facilitator collects and summarises participants' answers, which are then reported to them. In the subsequent round, the experts can be asked to revise their answers, or to answer more in-depth questions. The process terminates after a given number of rounds, or after convergence is reached. In its original formulation, a Delphi survey is a tool especially suitable for collective forecasting of a given phenomenon based on expert opinions.

For our study, we asked 10 social media and Web experts to answer a few questions designed after the main findings of our analysis. In spite of the spirit of a Delphi survey, the goal of this analysis is to collect thoughtful answers from a panel of experts, rather than to reach convergence. Briefly, we asked how the credibility of news can be affected by contextual hints, such as the source of an article, the informative richness of the article's preview on social networks, and what other users thought of the article; we also asked how often users rely on external fact-checking, and what age users are more likely to fact-check news. Round 1 questions were designed as open questions, which let participants point out their opinion with no constraints; Round 2 questions were designed as multiple choice questions, specifically targeted at the main topics and themes that emerged after Round 1. In the following, we report Round 1's questions:

\begin{itemize}
    \item Q1: \textbf{What are, based on your experience and competencies, the main indicators of the credibility of a news article on social networks?}
    
    Participants spontaneously converged on the source of the article (9 out of 10), followed at a distance by the sources cited by the article, the coverage of the news across diverse sources, and who is the social network users that broke the news.
    
    \item Q2: \textbf{In your experience, what makes you think a news piece could be false?}
    
    Participants pointed out mainly the poor quality of the news (8 out of 10), but also an unreliable publisher (5 out of 10) and the lack of support in other news outlets (3 out of 10).
    
    \item Q3: \textbf{Based on your experience, how do you rate the quality and quantity of information reported in the news previews on social networks?}
    
    Participants found the quality of news previews to be poor (7 out of 10), with a slight tendency to clickbait, regardless of the news publisher. Also quantity of information has been pointed out as insufficient (4 out of 10).
    
    \item Q4: \textbf{Does exist, according to you, a relationship between news credibility and the response of users on social networks? If you think it exists, could you explain what you think it is?}
    
    6 out of 10 participants believe that bold claims, especially from false news publishers, result in higher arousal among readers, which can also foster news' diffusion. 4 out of 10 participants did not acknowledge any direct relationship between credibility and public response.
    
    \item Q5: \textbf{How do you make sure a piece of news is credible, when you are in doubt, in your experience as a user?} 
    
    Participants unanimously indicated a parallel search on other news sources. Additionally, 4 out of 10 participants mentioned the need for checking on reliable sources, such as debunking websites.
\end{itemize}

From Round 1, it has emerged that the source of a news article can influence its credibility; that text in news previews can be insufficient; that roughly half of the experts think that users' response on social networks can be influenced by false, bold claims; that the most reasonable method for debunking a piece of news is to search it on other reliable sources. Round 2 questions, here reported, address the previous points with more specificity:

\begin{itemize}
    \item Q1 \textbf{Based on your knowledge of the domain, do you think that the source a news article is taken from is an important feature for assessing news credibility?}
    
    Most users agreed that news outlets play a role in the perception of credibility.
    
    \item Q2 \textbf{From Round 1 it emerged that information contained in news previews on social networks is often poor in terms of quality. Do you think that reading the entirety of an article can bring more useful information to decide whether the article is credible or not?}
    
    6 out of 10 users strongly agreed, while the remaining mildly agreed.
    
    \item Q3 \textbf{From the previous round, it emerged a relationship between the credibility of news and users' response. Based on your knowledge of the domain, do you think that an interface that shows the opinion of the majority of users about the credibility of the news piece can be effective in influencing the opinion of an individual?}
    
    Experts are divided, as 3 out of 10 answered "Not much", while 3 out of 10 answered "Very much".
    
    \item Q4 \textbf{From the previous round, it has emerged agreement about the need of checking the coverage of a news piece on other sources before deciding about its credibility. What do you think is the percentage of users that search for the credibility of some news before making up their minds on it?}
    
    4 out of 10 participants indicated "5\% to 10\%", while the remaining are equally divided into "10\% to 20\%" and "20\% to 30\%".
    
    \item Q5 \textbf{Considering a given percentage of users that check for the credibility of news on other sources, do you think that the majority of them belong to one (or more) of which of the following generations?}
    
    60\% of answers indicated "Millennials 1981-96" as the generation more likely to fact-check news online, while only a minority indicated older generations, such as "Gen X (1965-80)", or younger, such as "Gen Z (1997-2012)" (both at 20\%). For almost one-third of respondents there is no relationship between age and the propensity to search for the same news on different outlets.

\end{itemize}

Round 2 answers are, to a small extent, in contrast with the quantitative findings of our analysis. The source of news was acknowledged as a very informative feature, confirming our findings (see Section~\ref{sec:results}). Respondents to the survey agreed that reading an article in its entirety could help determine its credibility; however, results reported in Section~\ref{subsec:full_text} show that this may be a misconception: users that took the Fakenewslab test performed worse, in terms of accuracy, when confronted with long texts, possibly due to the fatigue of reading excessive information. The survey panellists are divided about the effectiveness of peer pressure on individual judgement, a behaviour well-documented in literature and further confirmed by our analysis. They also underestimated the percentage of users that would search for an article on other venues, that in our test was assessed around 18\% - it must be reminded, though, that it is impossible to determine if users that left the browser's tab and came back later did actually fact-checked the news: this finding must be taken carefully. Last, users correctly indicated young adults as the most likely to carry out parallel research of the same article on other news; one-third of respondents, however, did not acknowledge any relationship between age and propensity to fact check.

Coupled with the quantitative analysis we discussed in Section~\ref{sec:results}, this qualitative feedback shed light on the most common misconceptions about how users assess the credibility of news articles online, revealing some interesting asymmetries between what experts think is important and what users actually considered to be important, when taking the Fakenewslab test.

\newpage

\newpage
\bibliographystyle{abbrev-doi}
\bibliography{bib}

\end{document}